\documentclass[a4paper,11pt]{article}
\usepackage{jinstpub}

\usepackage{here}
\usepackage[dvipsnames]{xcolor}
\usepackage[T1]{fontenc}
\usepackage{tgcursor}
\usepackage{verbatim}

\title{Interstrip Capacitances of the Readout Board used in Large Triple-GEM Detectors for the CMS Muon Upgrade}

\author[n]{M.~Abbas,} 
\author[t]{M.~Abbrescia,}
\author[h,j]{H.~Abdalla,}
\author[h,k]{A.~Abdelalim,}
\author[h,i]{S.~AbuZeid,}
\author[d]{A.~Agapitos,}
\author[ae]{A.~Ahmad,}
\author[q]{A.~Ahmed,}
\author[ae]{W.~Ahmed,}
\author[y]{C.~Aim\`e,}

% ...
\affiliation[d]{Peking University, Beijing, China} %
% ...
\affiliation[h]{Academy of Scientific Research and Technology - ENHEP, Cairo, Egypt} %
\affiliation[i]{Ain Shams University, Cairo, Egypt}
\affiliation[j]{Cairo University, Cairo, Egypt}
\affiliation[k]{Helwan University, also at Zewail City of Science and Technology, Cairo, Egypt}
% ...
\affiliation[n]{Karlsruhe Institute of Technology, Karlsruhe, Germany}
% ...
\affiliation[q]{Delhi University, Delhi, India}
% ...
\affiliation[t]{Politecnico di Bari, Universit\`{a} di Bari and INFN Sezione di Bari, Bari, Italy}%
% ...
\affiliation[y]{Universit\`{a} di Pavia and INFN Sezione di Pavia, Pavia, Italy} %
% ...
\affiliation[ae]{National Center for Physics, Islamabad, Pakistan}
% ...

\emailAdd{stephen.butalla@cern.ch}
\emailAdd{shivali.malhotra@cern.ch}

\abstract{We present analytical calculations, Finite Element Analysis modeling, and physical measurements of the interstrip capacitances for different potential strip geometries and dimensions of the readout boards for the GE2/1 triple-Gas Electron Multiplier detector in the CMS muon system upgrade. The main goal of the study is to find configurations that minimize the interstrip capacitances and consequently maximize the signal-to-noise ratio for the detector. We find agreement at the 1.5--4.8\% level between the two methods of calculations and on the average at the 17\% level between calculations and measurements. A configuration with halved strip lengths and doubled strip widths results in a measured 27--29$\%$ reduction over the original configuration while leaving the total number of strips unchanged. We have now adopted this design modification for all eight module types of the GE2/1 detector and will produce the final detector with this new strip design.}

\keywords{Detector modeling and simulations II;
Micropattern gaseous detectors
}

\begin{document}
\maketitle
\flushbottom

\section{Introduction}\label{sec:int}
The Large Hadron Collider (LHC) was built to shed light on several fundamental questions in particle physics. The Compact Muon Solenoid (CMS) experiment \cite{CMS} is one of the LHC's two general purpose experiments designed and built to detect and reconstruct particles produced in proton-proton (pp) and heavy ion (proton-ion and ion-ion) collisions. With the discovery of the Higgs boson \cite{HiggsATLAS, HiggsCMS}, the CMS Collaboration has established a rigorous research program involving precise measurements of Higgs boson properties and searches for new physics. This requires a large increase in the LHC luminosity, which puts stringent requirements on the detectors. In order to maintain its excellent performance, the CMS experiment is currently undergoing a series of upgrades of its components, including its muon system \cite{CMSTDR,MuonTDR}. The upgrade of the muon system is a critical component of CMS due to the strong role of exploring new physics with  muons in the final state.

\begin{figure}[!htp] 
\centering
\includegraphics[width=0.9\textwidth]{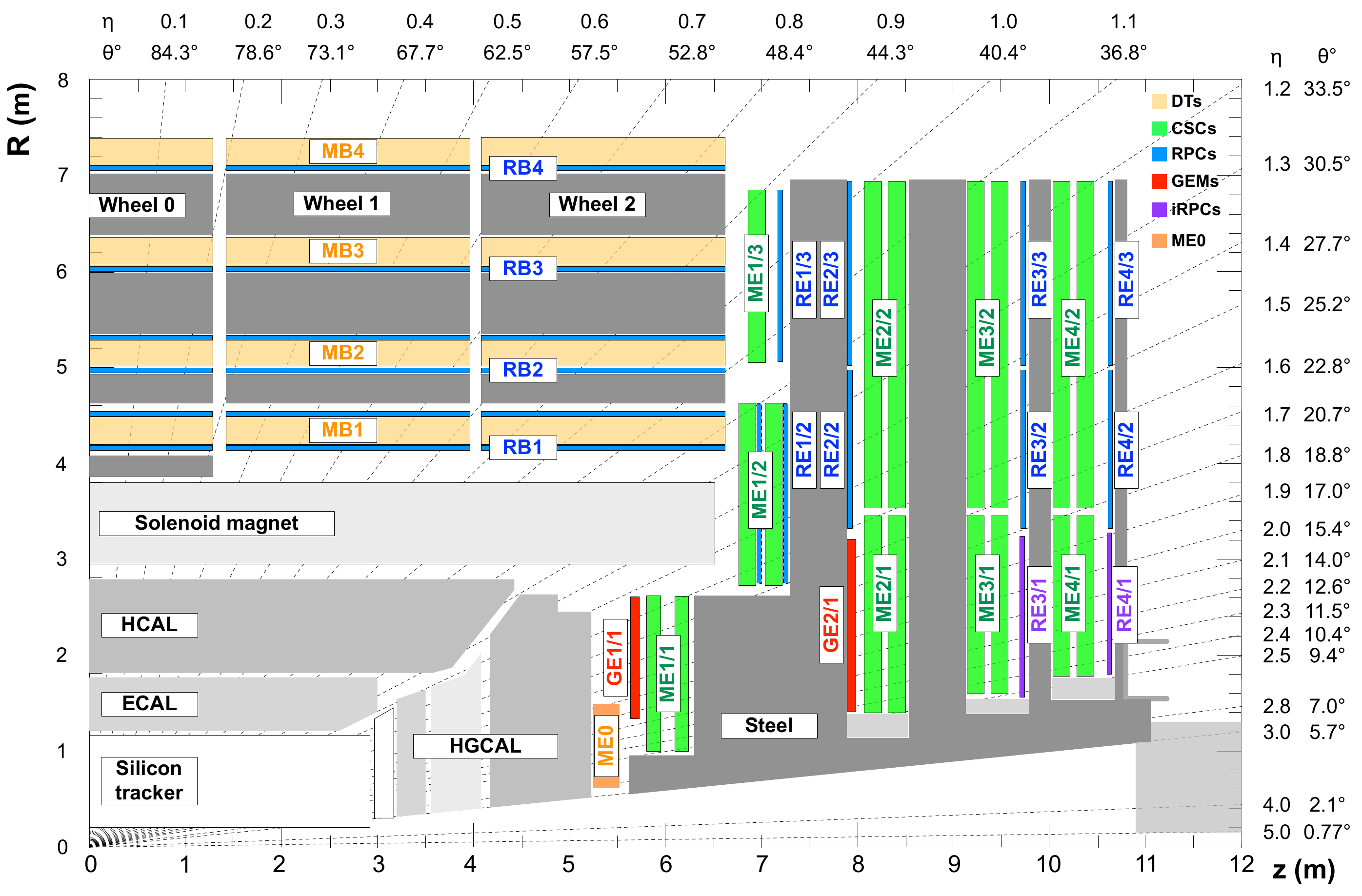}
\caption{\label{fig:endcap} An R-z cross section of a quadrant of the upgraded CMS detector high-lighting the locations of the new GE1/1, GE2/1, and ME0 stations with GEM technology in the CMS muon endcap region. The previously existing muon stations, i.e.\ drift tubes (MB), cathode strip chambers (ME), and resistive plate chambers (RB, RE),  and the flux-return steel yoke (dark areas) are also shown.}
\end{figure}

\begin{figure}[!h] 
\centering
\includegraphics[width=0.63\textwidth]{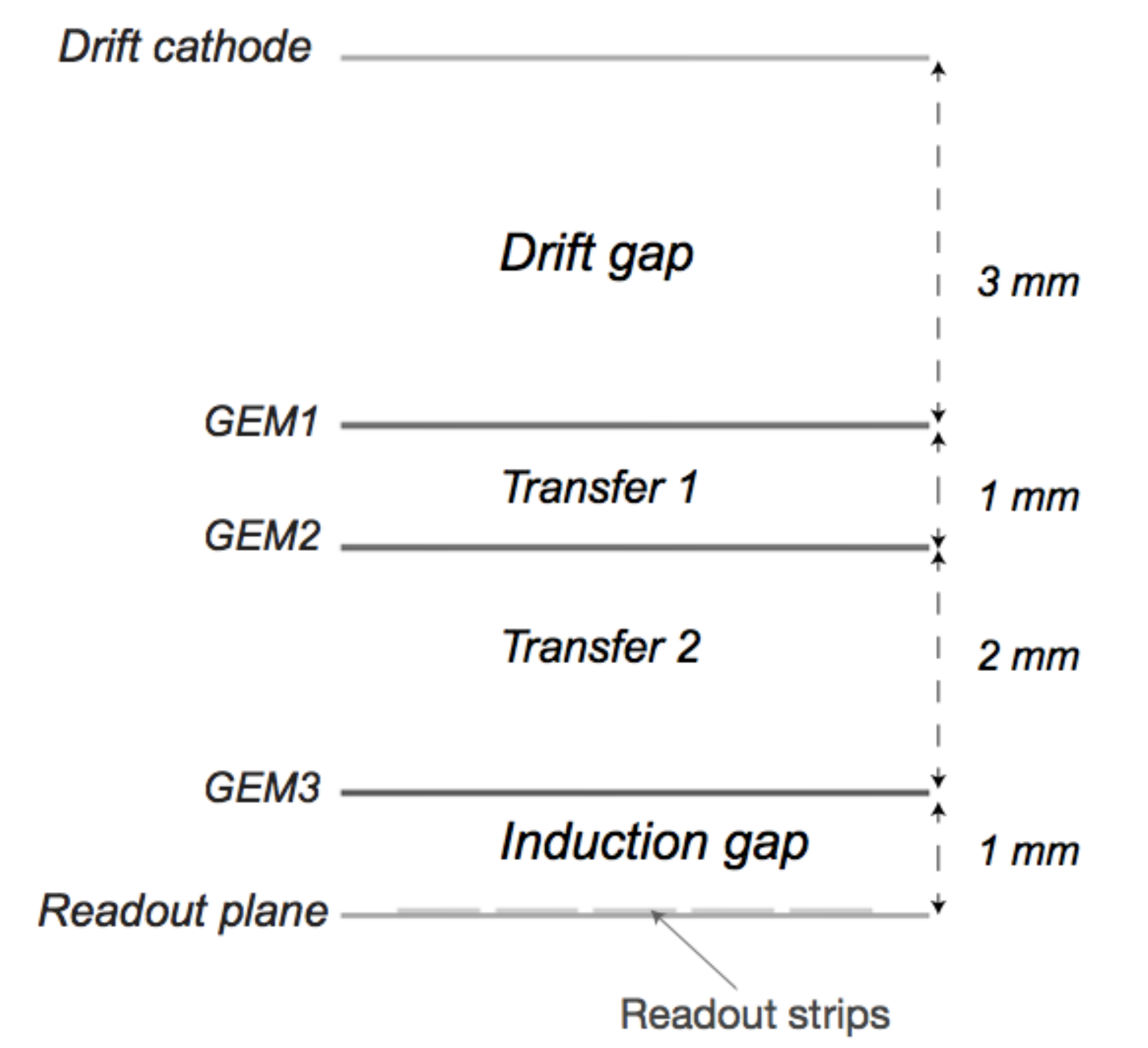}
\caption{\label{fig:Triple-GEM}Cross-sectional view of a CMS triple-GEM detector consisting of three GEM foils.}
\end{figure}

The CMS muon system is composed of three detector technologies: Resistive Plate Chambers (RPC), Drift Tubes (DT) and Cathode Strip Chambers (CSC), all of which are being upgraded \cite{MuonTDR}. To ensure continued function of the muon trigger at acceptably low Level-1 trigger rates and to increase redundancy and acceptance of the muon system, a new muon subdetector based on Gas Electron Multipliers (GEMs) \cite{Sauli} is being added in the forward region of the CMS detector (see figure \ref{fig:endcap}) \cite{MuonTDR, GEMTDR}. For this upgrade, triple-GEM detectors, i.e.\ micro-pattern gas detectors with three GEM foils, are being used. Figure~\ref{fig:Triple-GEM} shows a schematic cross section of the geometrical configuration of drift anode, foils, and readout board for all CMS triple-GEM detectors.
\newpage

\newpage

One of the challenges with any detector system is improving the signal-to-noise ratio (S/N). One variable that influences the noise in the detector is the capacitance between readout strips on the readout board (ROB),  that we refer to as the ``interstrip capacitance''. As the S/N is influenced by the geometrical configuration and the dimensions of the readout strips, i.e.\ the length and width of the strips, as well as their spacing, the interstrip capacitance is crucial for detector operation and performance. The purpose of this study is to optimize the final readout strip geometry of the GEM detectors in the GE2/1 station (see figure~\ref{fig:endcap}) to maximize the S/N. 

Specifically, we are addressing the concern that the strips are quite long in the original design of the largest GE2/1 modules. This leads to significant capacitances presented to the inputs of the front-end electronics and a danger of unacceptable noise levels. We consider modifying the original design~\cite{MuonTDR} by cutting the strips in half and doubling their widths while keeping the gap between strips the same. This obviously preserves the area of each strip and consequently the total number of strips per module. The key question then is by how much exactly this changes the interstrip capacitance. This motivates the studies presented here.

We discuss results from three methods used to determine the interstrip capacitance: analytical calculation, two- and three-dimensional (2D and 3D, respectively) simulations using Finite Element Analysis (FEA), and experimental measurements on a custom ROB with different strip geometries. 

 This paper is organized as follows: In section~\ref{sec:geometry}, we briefly describe the overall geometry of the GE2/1 chambers and of the readout strips on the readout boards. Section~\ref{sec:calc} provides information on the calculation and modeling of the interstrip capacitance of the GE2/1 strips. In section~\ref{sec:measure}, we describe the setup used for experimental measurements of the interstrip capacitances. Results are presented and discussed in section~\ref{sec:results}.

\section{Geometry of CMS GE2/1 GEM detectors}
\label{sec:geometry}
\begin{figure}[!htp] 
\centering
\includegraphics[width=0.45\textwidth,trim = {0.2cm 0 0 0 },clip]{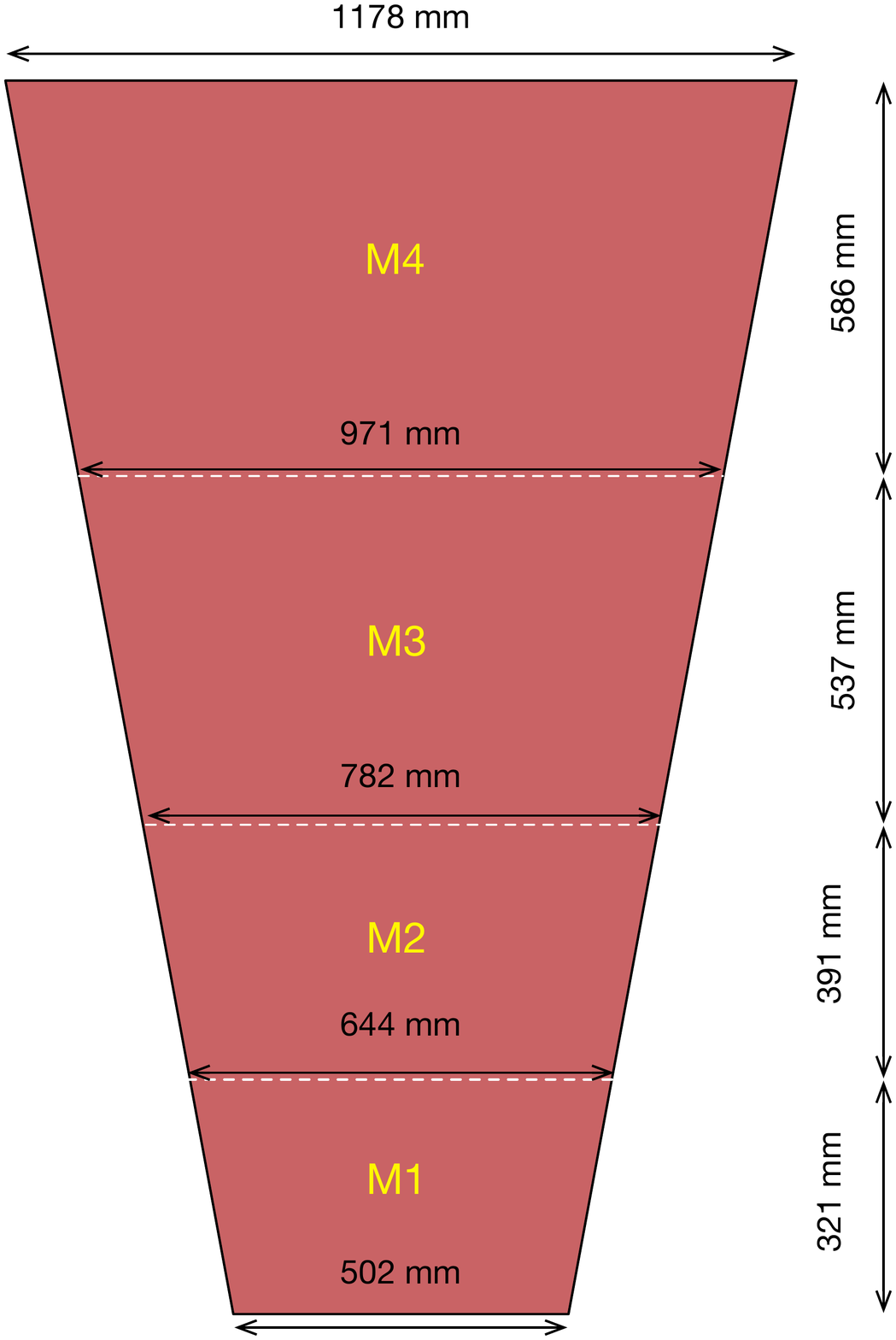}\hspace{0.05\textwidth}
\includegraphics[width=0.45\textwidth,height=0.42\textheight]{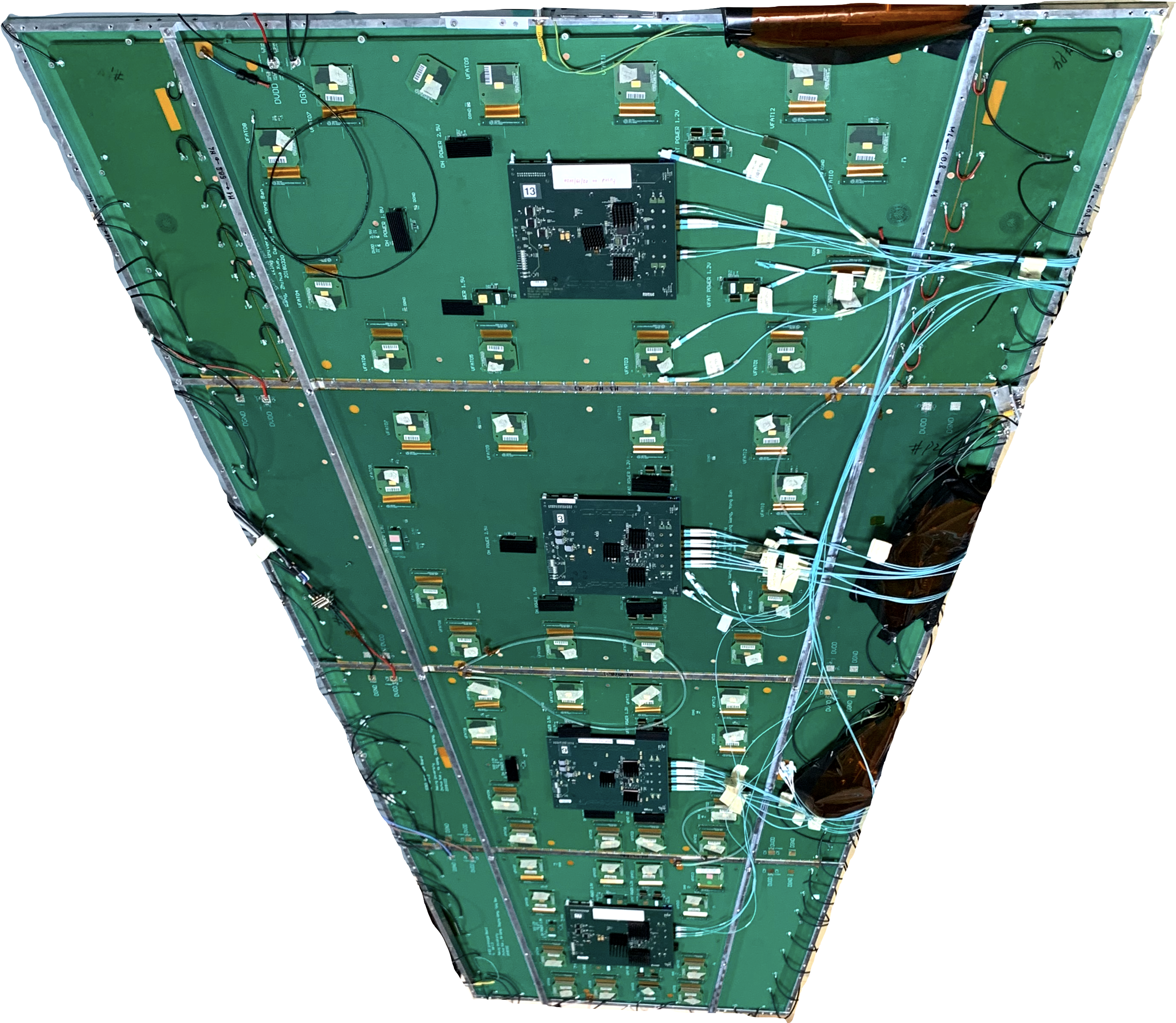}
\caption{\label{fig:GE21}Left: The dimensions of the modules in the back chamber of the GE2/1 superchamber. Right: A fully instrumented GE2/1 prototype chamber.}
\end{figure}

The GE2/1 muon station will cover the pseudo-rapidity region $1.6 < |\eta| < 2.4 $. This station consists of large trapezoidal chambers as shown in figure~\ref{fig:GE21}. A GE2/1 chamber comprises four separate modules and covers an area of 1.45 m$^{2}$. Chambers are installed in pairs on the muon station to provide two positional measurements per muon track. The GE2/1 station will comprise 72 chambers, each covering 20.3$^{\circ}$ in the azimuthal direction, with 36 chambers per muon endcap. The back chamber contains four modules labeled M1 (smallest) to M4 (largest), and the front chamber comprises corresponding modules M5 (smallest) to M8 (largest). In total, 864 large GEM foils are needed for this system. In this study, the smallest and largest modules of a GE2/1 chamber, i.e.\ the M1 and M4 modules, are considered; their original strip specifications are summarized in table~\ref{tab:GE21modules}~\cite{MuonTDR}.
\newpage 

\begin{table}[!htp]
\centering
\caption{\label{tab:GE21modules}Original specifications of readout strips used in M1 and M4 modules.}
\smallskip
\begin{tabular}{|cccc|}
\hline
Module & Strip length (cm) & Strip width (cm) & Interstrip gap (cm)\\
\hline
M1 & 19.6 & 0.0478 & 0.02\\
M4 & 20.6 & 0.124  & 0.02\\
\hline
\end{tabular}
\end{table}

\begin{figure}[!htp] 
\centering
\includegraphics[width=0.95\columnwidth]{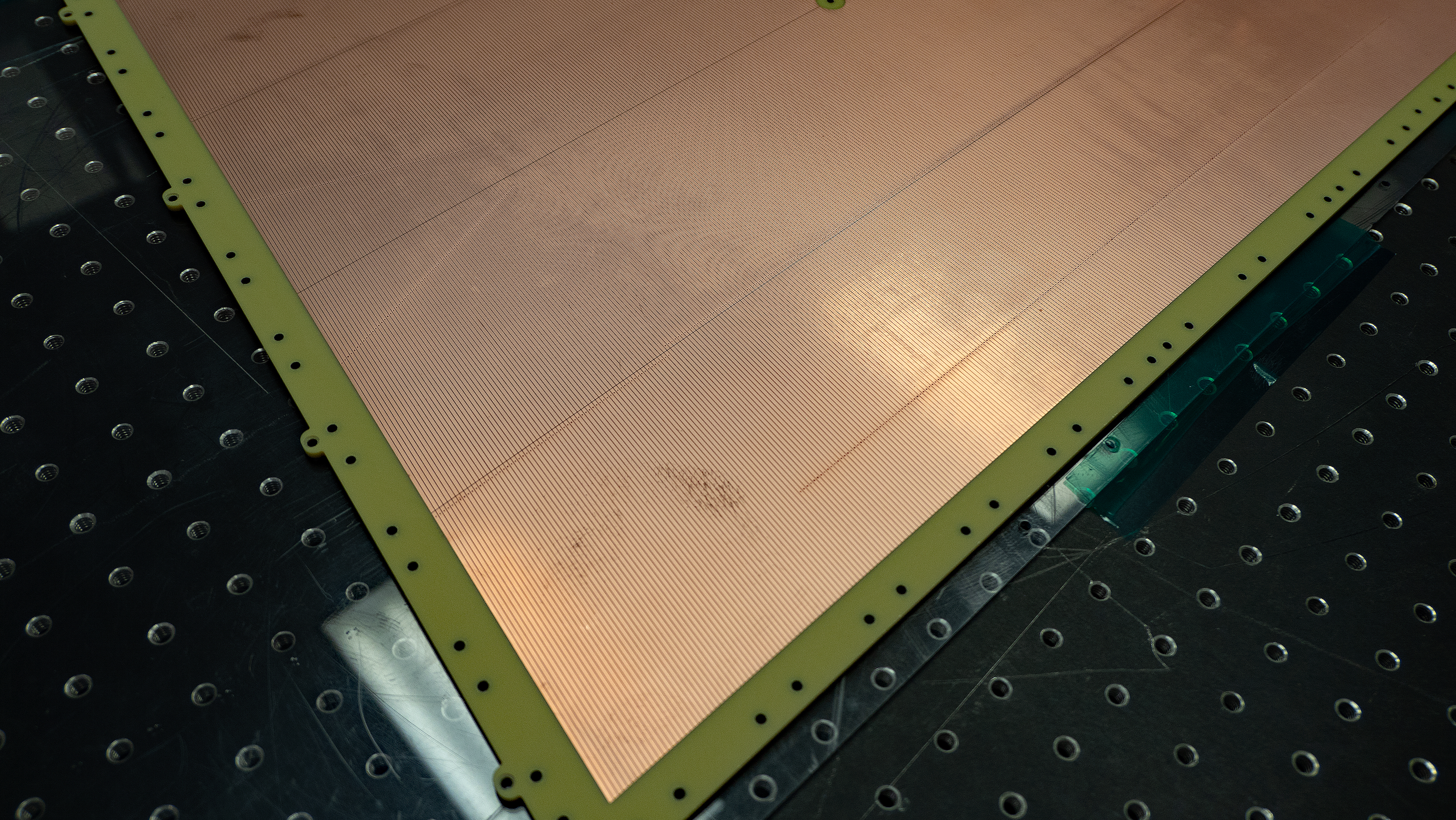}\\
\includegraphics[width=0.95\columnwidth]{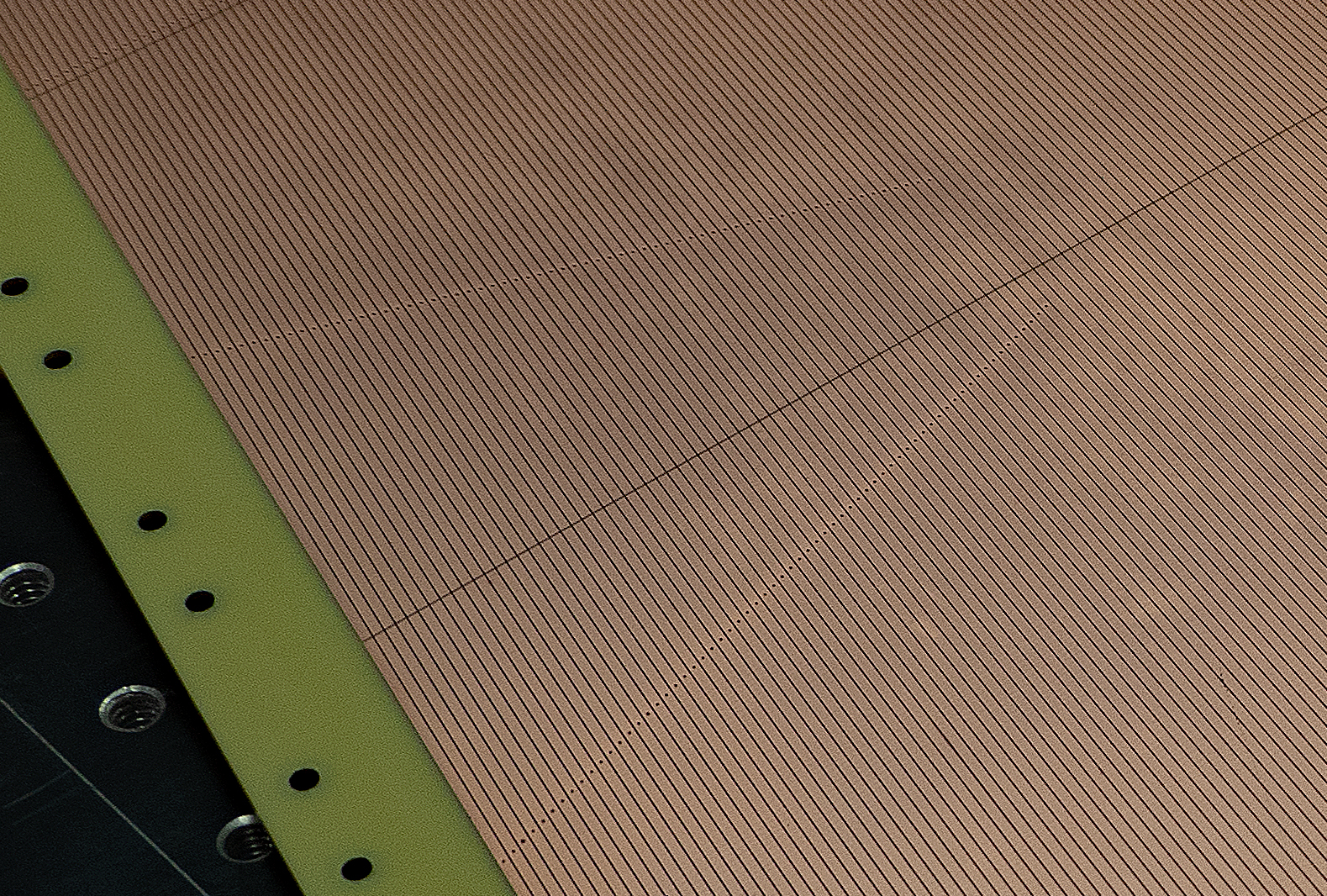}
\caption{\label{fig:M5ROB}Top: The readout strips on the ROB of a prototype module of the GE2/1 GEM detector. Bottom: A zoomed image of readout strips in the same ROB. Note the vias running in diagonal patterns near the center.}
\end{figure}

\noindent Figure~\ref{fig:M5ROB} (top) shows a picture of the readout strips of the readout board of an M5 module as an example. The bottom of the same figure shows a zoomed image of the readout strips. Note the readout strip segmentation between readout sectors, as well as the diagonal pattern of vias that connect the strips to connecting traces on the back of the printed circuit board (PCB).
\newpage

\section{Calculation and modeling}
\label{sec:calc}
\begin{figure}[!htp] 
\centering
\includegraphics[width=0.7\textwidth]{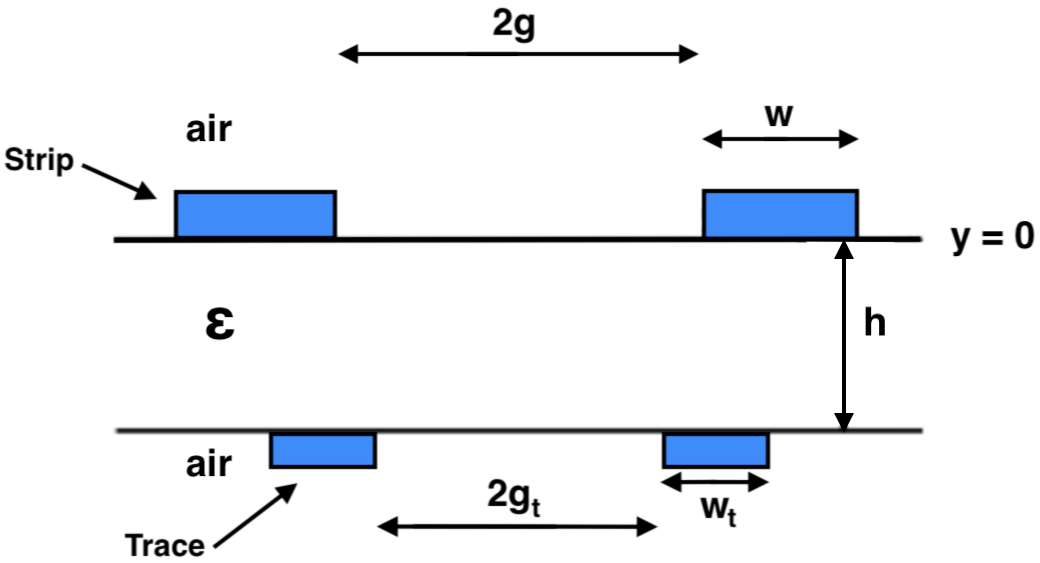}
\caption{\label{fig:robsection}The readout strip and trace geometry, where w is the strip width, $2g$ is the gap width of the strips, $w_{t}$ is the trace width, $2g_{t}$ is the gap width of the traces, and h is the thickness of the substrate with dielectric constant $\epsilon$. Note that strips and traces are connected by vias (not shown), which are conductive connections that run between the layers of the PCB.}
\end{figure}

Figure~\ref{fig:robsection} displays a cross section of a subsection of the GE2/1 ROB used for the analytical calculations. Here, ``Strip'' refers to the readout strips on the ROB and ``Trace'' refers to the signal lines on the other side of the ROB which are connected to the readout strips through a via, which is an electrically conductive channel that runs through the PCB. These signal traces terminate in the readout connectors, where the front-end electronics are plugged in so that a signal can be read out.

\subsection{Analytical calculation}

The capacitance between coplanar metal strips on a dielectric surface has been widely investigated and used in particular in the fields of telecommunication and microwave applications \cite{Capa1, Capa2}. In this work, the interstrip capacitance per centimeter of strip length is calculated using a modified version of the expression developed in~\cite{Capa1}. This equation is modified to account for readout strips on the front of the board and for the traces on the back side of the ROB. Using the dimensions displayed in figure~\ref{fig:robsection}, we obtain the following equations, which are a linear sum of the strip and trace capacitances:
\begin{align} \label{1}
(C/l)&=(C_{sa}+C_{a})l_{s}^{-1}+(C_{st}+C_{t})l_{t}^{-1}\\
&= \epsilon_{0}\left( (\epsilon-1)\frac{K(k')}{2K(k)}+\frac{K(k_{0}^{'})}{K(k_{0})} \right) \nonumber\\
&+ \epsilon_{0}\left( (\epsilon-1)\frac{K(k'_{t})}{2K(k_{t})}+\frac{K(k_{0t}^{'})}{K(k_{0t})} \right)\nonumber
\end{align}
where:
\begin{align} \label{2}
k&=\tanh\left(\frac{\pi g}{2h}\right)\coth\left(\frac{\pi(w+g)}{2h}\right)\\
k' & =\sqrt{(1-k^2)}\\
k_{0}&=\frac{g}{w+g}\\
k_{0}^{'} & =\sqrt{(1-k_{0}^2)}\\
k_{t}&=\tanh\left(\frac{\pi g_{t}}{2h}\right)\coth\left(\frac{\pi (w_t +g_{t})}{2h}\right)\\
k^{'}_t & =\sqrt{(1-k_{t}^2)}\\
k_{0t} & =\frac{g_{t}}{w_{t}+g_{t}}\\
k_{0t}^{'} &=\sqrt{(1-k_{0t}^2)}
\end{align}

\noindent Here, $C_{a}$ is the capacitance between two readout strips of length $l_{s}$ with air above and below, $C_{sa}$ is the capacitance between the readout strips due to the presence of the PCB substrate below, $C_{t}$ is the interstrip capacitance between two traces of length $l_{t}$ with air above and below, and $C_{st}$ is the capacitance between the traces due to the presence of the substrate.
Equations 3.2-3.9 are the moduli of $K(k)$, which is the complete elliptic integral of the first kind, where $w$ is the strip width, $w_{t}$ is the trace width, 2$g$ is the gap between the readout strips, 2$g_{t}$ is the gap between the traces, $h$ is the thickness of the FR4 (flame retardant glass-reinforced epoxy laminate) substrate with dielectric constant $\epsilon =4.7$, and $\epsilon_{0}$ is the vacuum permittivity.

The average trace lengths, trace widths, and gap widths used for this calculation are measured experimentally as explained in section~\ref{sec:measure}. The calculations are performed using {\fontfamily{qcr}\selectfont MATLAB}~\cite{MATLAB}, and the built-in \texttt{MATLAB} function {\fontfamily{qcr}\selectfont
ellipke()} is used for the complete elliptic integral of the first kind.

\subsection{Finite Element Analysis model}

\begin{figure}[!htp] 
\centering
\includegraphics[width=0.44\textwidth]{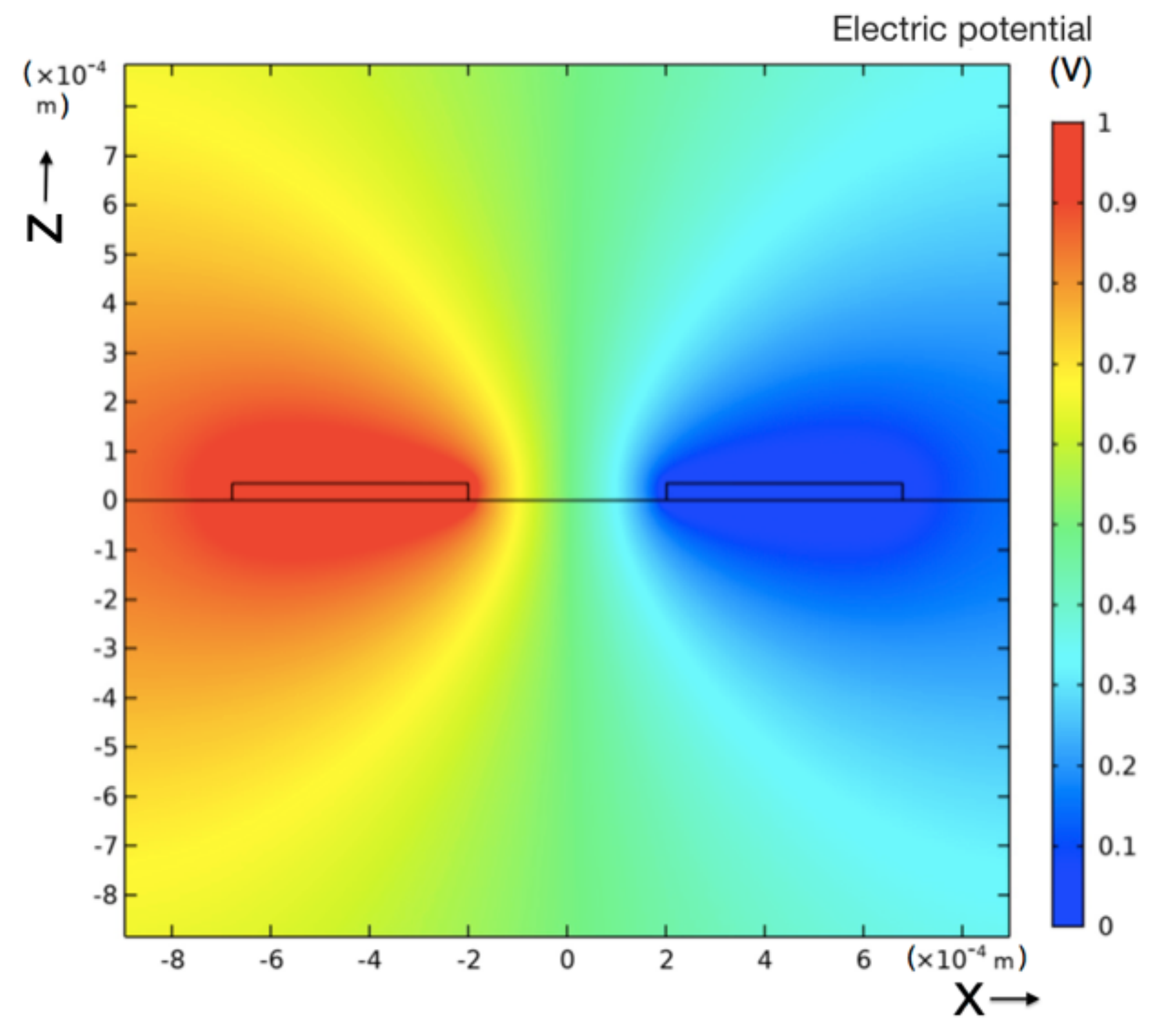}
\includegraphics[width=.5\textwidth]{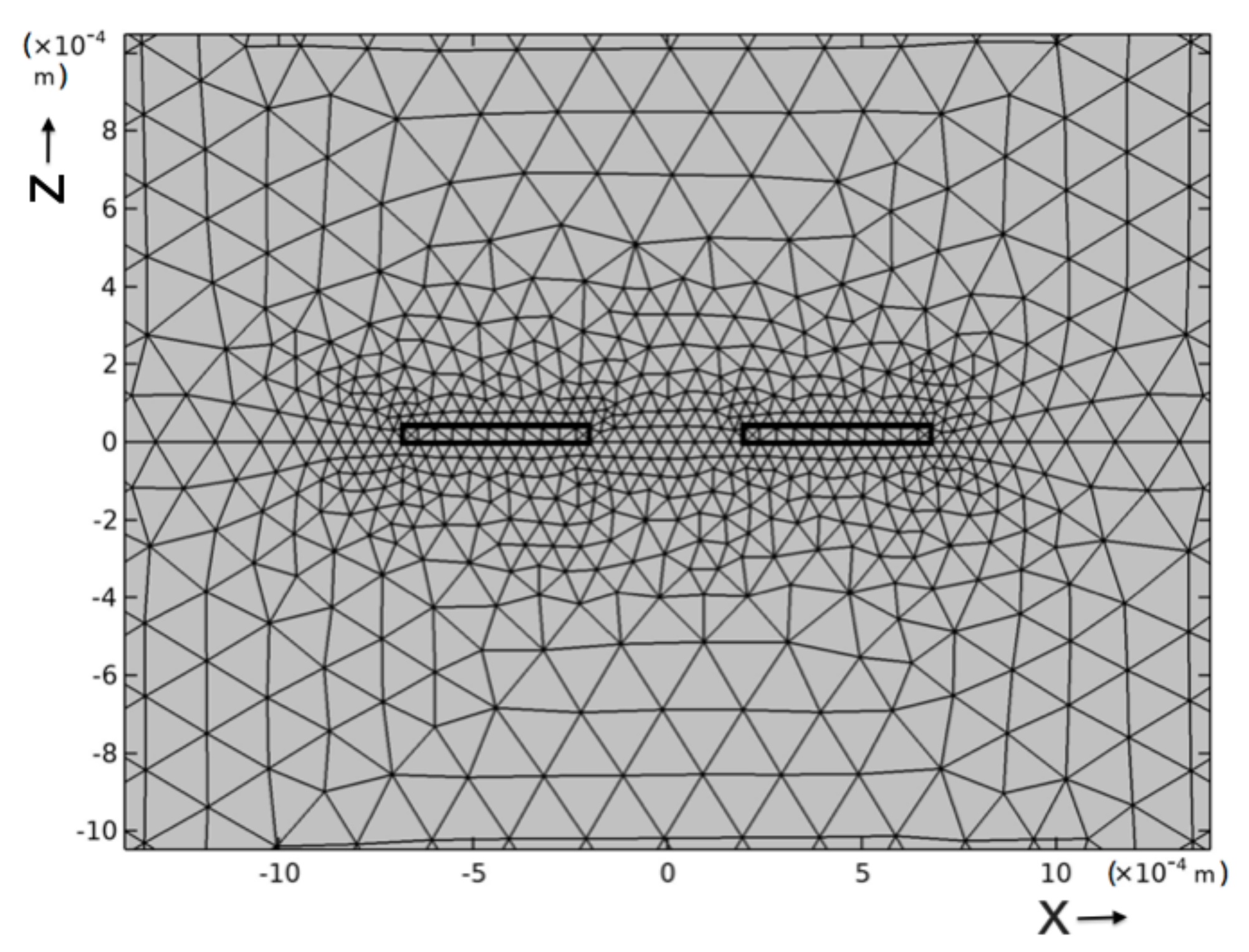}
\caption{\label{fig:2_strips}Left: The electric potential V in the vicinity of two readout strips with a strip gap of 0.4 mm calculated with the 2D FEA model. A potential difference of 1 V is applied between the strips and the interstrip capacitance is found to be 0.445 pF/cm. Right: The structure of the mesh obtained in the strips along with the air volume on the top and substrate volume at the bottom of the strips.}
\end{figure}

A model of the GEM readout board based on the FEA is created using \texttt{COMSOL}, a Multiphysics simulation software \cite{COMSOL}. The initial model uses simply two strips with the strip width and the width of the gap between the two strips as parameters. This results in a basic 2D model, as shown in figure~\ref{fig:2_strips} (left). To find the interstrip capacitance, a potential difference of 1~V is applied between the two strips and then the COMSOL software calculates the charges $q$ using Gauss's Law and the capacitance $C=\frac{q}{V}$. Specifically, the FEA model always considers one strip at 1~V and the other one at 0~V. The model utilizes the Electrostatic Physics module and extra fine meshing is done for all the components as shown in figure~\ref{fig:2_strips} (right). 

\begin{figure}[!htp] 
\centering
\includegraphics[width=.41\textwidth]{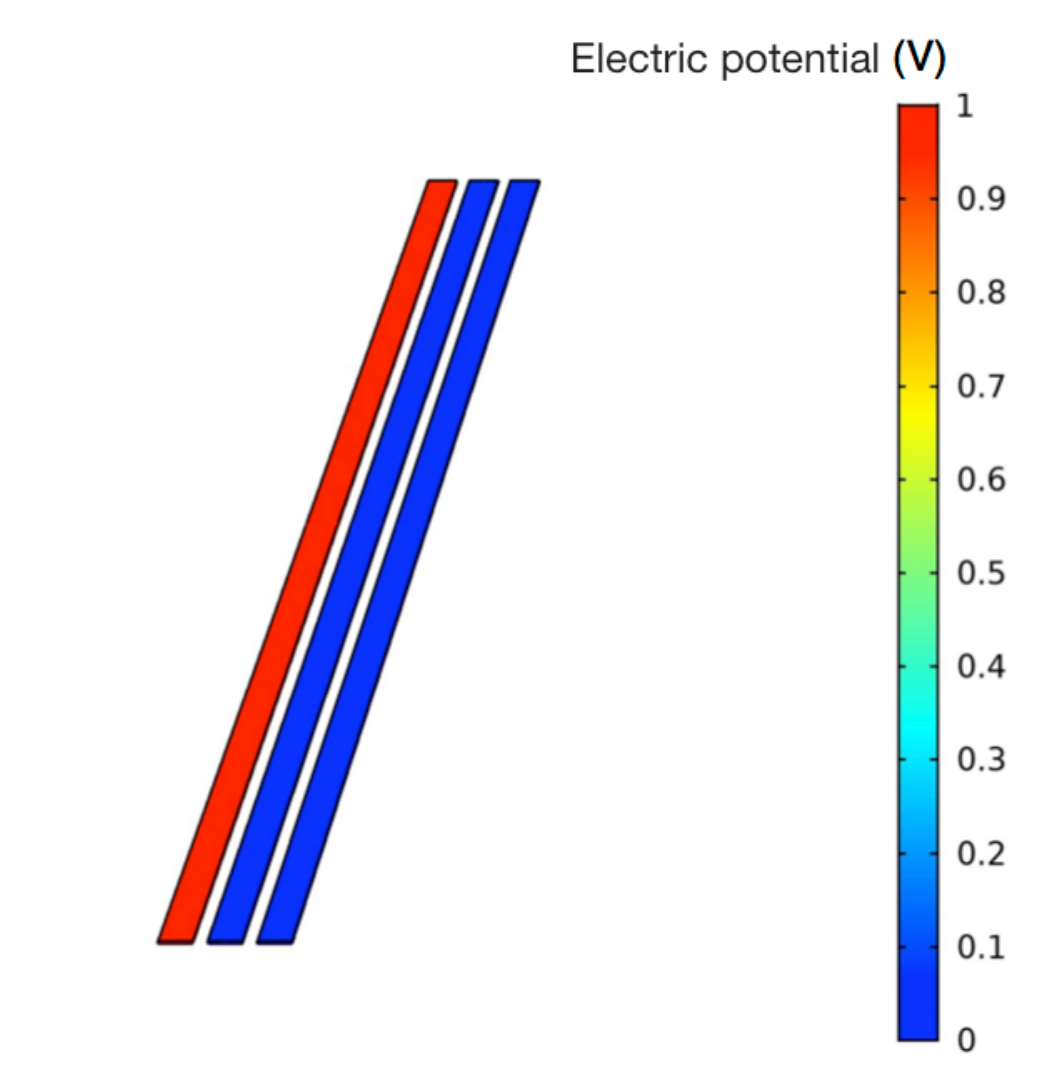}
\hspace{0.1cm}
\includegraphics[width=.47\textwidth]{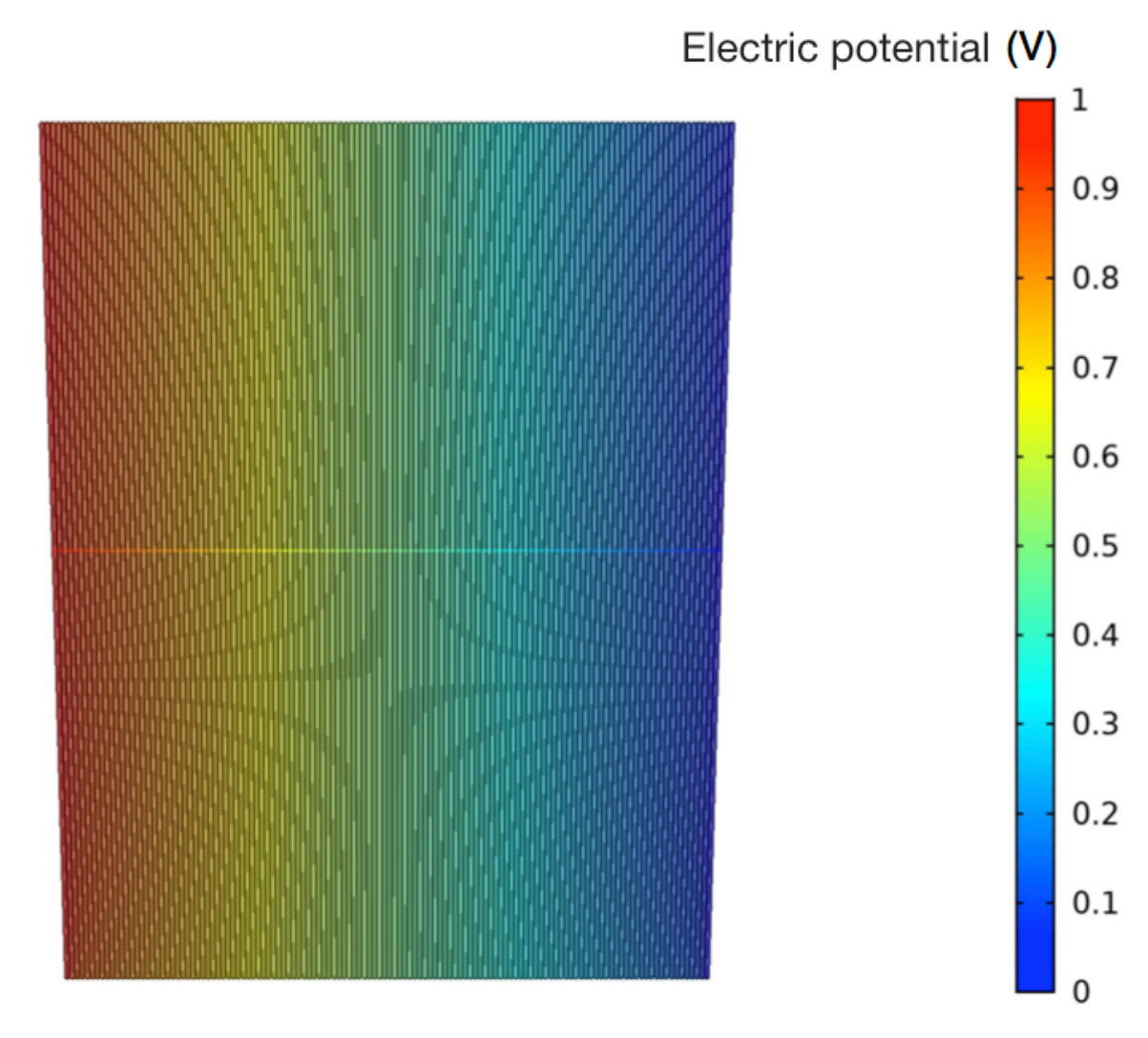}
\caption{\label{fig:3_128_strips}Left: 3D model showing different electric potential between three strips. Right: 3D model showing different electric potential between 128 strips. (Note: Substrate is present below the strips in both cases but for clarity it is omitted in the images shown above.)}
\end{figure}

\begin{figure}[!htp] 
\centering
\includegraphics[width=0.45\columnwidth]{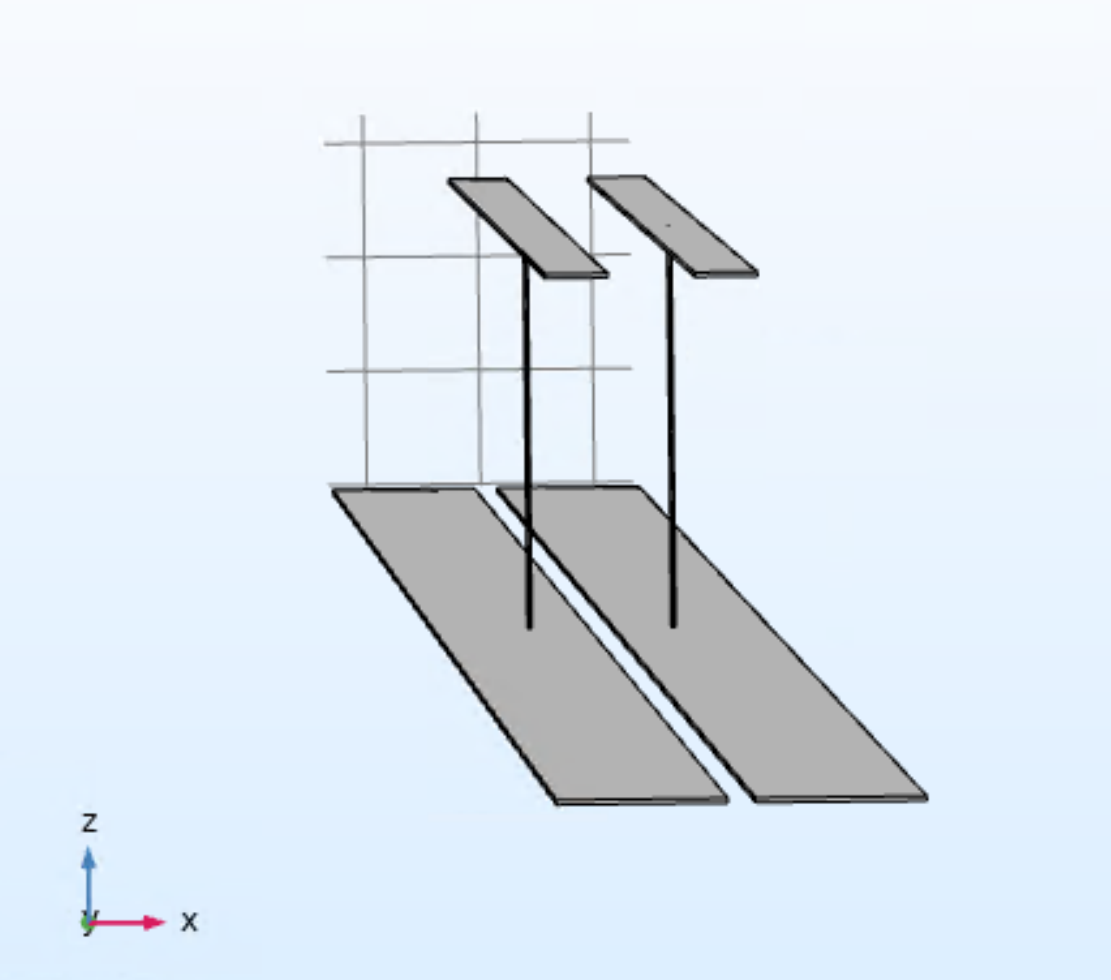}
\includegraphics[width=0.41\columnwidth]{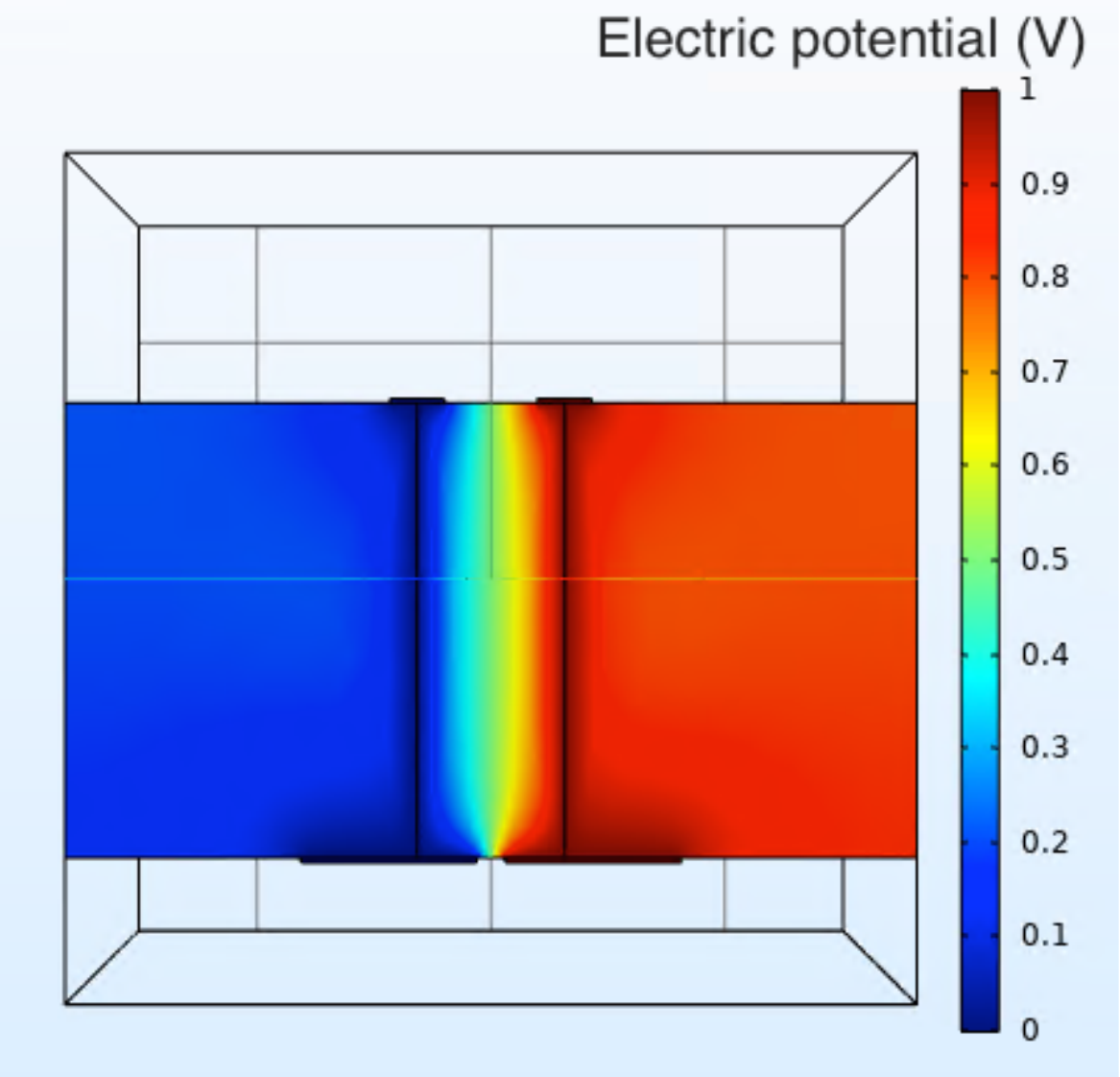}
\caption{\label{fig:strips_traces}Left: The 3D model with two readout strips at the bottom connected by vias to traces at the top. Right: The resulting electric potential map after applying a 1~V potential difference between the two traces. (Note: For clarity, the left image does not show the presence of the substrate and the air volume even though they are taken into account in the calculation of the shown potential).}
\end{figure}

The analytical method is limited to two strips and 2D calculations. In order to get more detailed insight into the interstrip capacitance, we also perform 3D modeling using the \texttt{COMSOL} software, i.e.\ including the finite length of the readout strips and the 3D geometry. Figure~\ref{fig:3_128_strips} shows 3D models for three strips and for 128 strips. The latter can be utilized to calculate the interstrip capacitance between any strips, i.e.\ not limited to only adjacent strips. The 3D model with two strips can also be extended to take into account the traces along with the vias that connect them to the readout strips, as shown in figure~\ref{fig:strips_traces} (left). Here, the vias are connected from the center of the traces to the center of the strips. Figure~\ref{fig:strips_traces} (right) shows the resulting potential in cross-sectional view at the strip center.

\begin{figure}[!htp] 
\centering
\includegraphics[width=.49\textwidth]{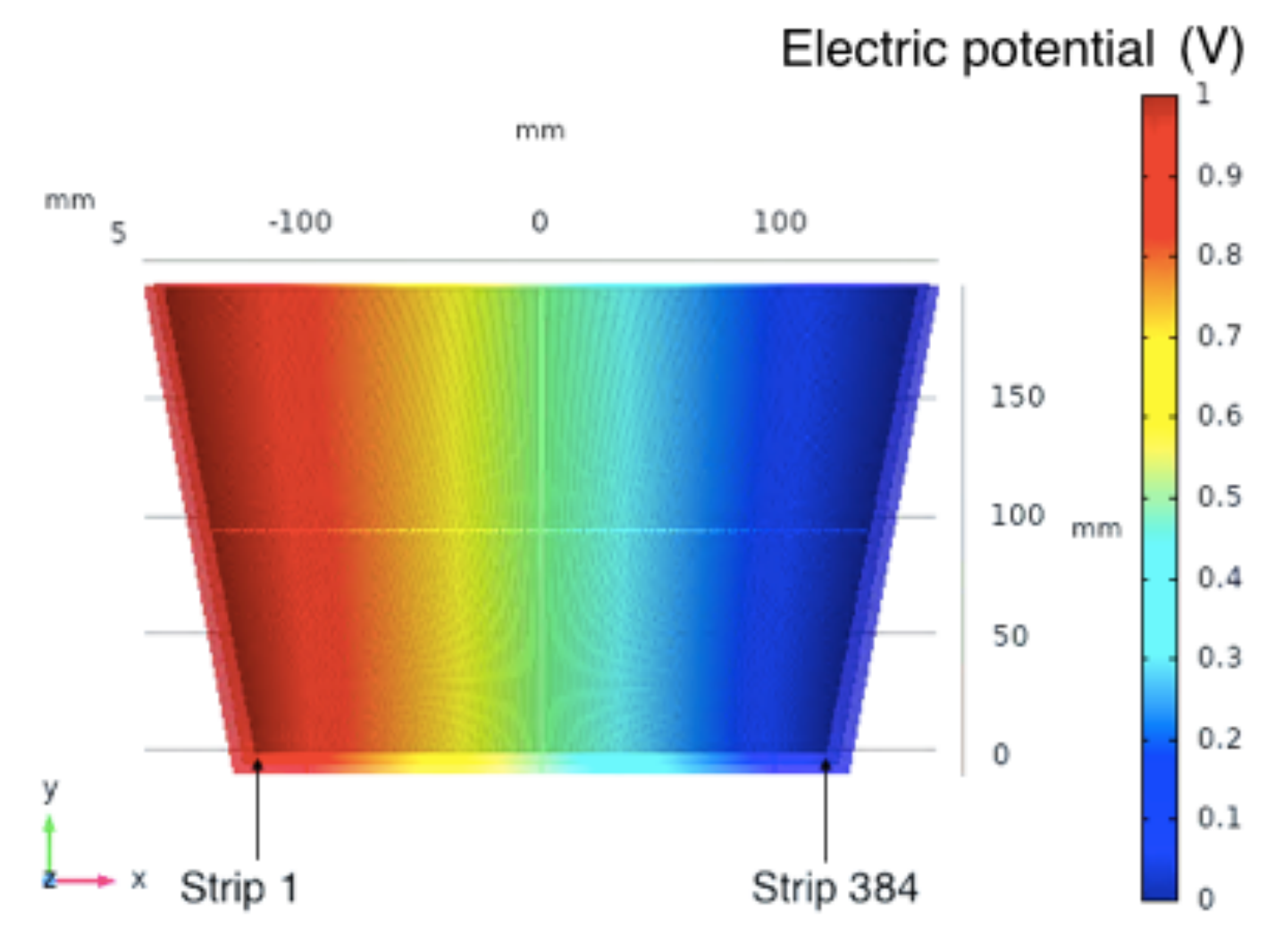}
\includegraphics[width=.5\textwidth]{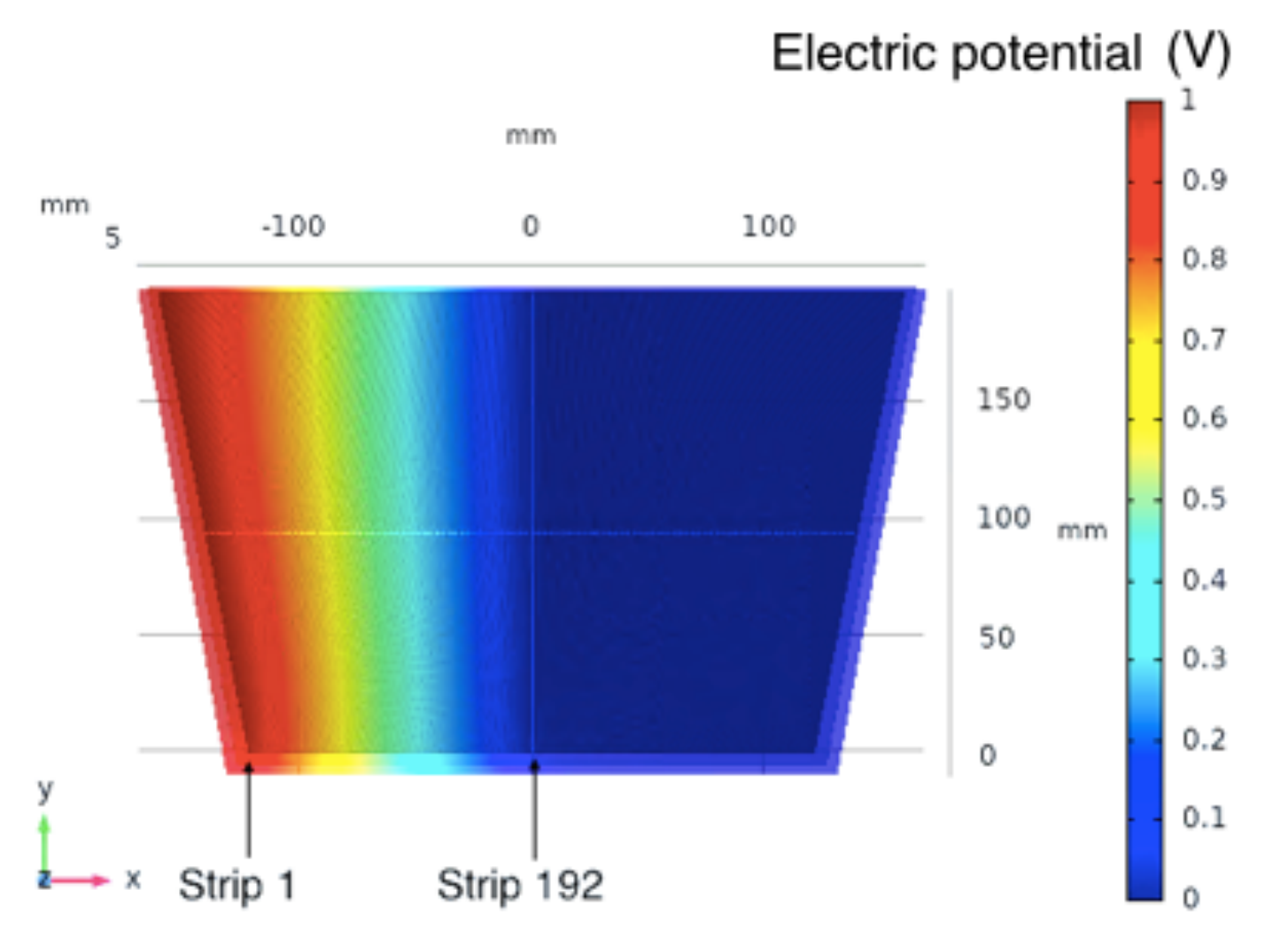}
\caption{\label{fig:M1_strips}Left: Electric potential across the bottom half of the M1 module for calculating the interstrip capacitance between first and  384\textsuperscript{th} strip. Right: Electric potential across the bottom half of the M1 module for calculating the interstrip capacitance between first and 192\textsuperscript{nd} strip. A fixed potential of 1~V (red) is applied to the first strip and a fixed potential of 0~V (blue) is applied to the other strip of interest. The capacitance between those two strips is then calculated with the potential of all other strips varying accordingly.}
\end{figure}

The models for the M1 and M4 modules of the GE2/1 chamber are also built using the \texttt{COMSOL} software. Since these are multi-strip models, it is not feasible to manually create the whole model. Instead, the \texttt{COMSOL} software is interfaced with \texttt{MATLAB} programming code to generate a model with the full complement of 384 strips. In this model, we also accommodate the gas volume in addition to the copper strips and the FR4 substrate. This model can also accommodate a copper sheet 1 mm above the ROB to simulate the presence of the GEM3 foil above the ROB (see figure~\ref{fig:Triple-GEM}), consistent with the hardware configuration (see figure~\ref{fig:ROBwPCB}). 
Using this software, we can designate the strips between which we wish to calculate the interstrip capacitance.
This allows us to obtain the electric potential across the entire module and to calculate the interstrip capacitance between various combinations of the strips. For example, figure~\ref{fig:M1_strips} shows the electric potential across the M1 module with the value of the interstrip capacitance between first and 384\textsuperscript{th} strip determined to be 0.00838 pF/cm (left) and 0.0154 pF/cm between the first and the 192\textsuperscript{nd} strip (right). 

\section{Experimental measurements}
\label{sec:measure}
In order to conduct a comprehensive analysis of the problem, a custom ROB is fabricated that allows direct physical capacitance measurement. It has the overall shape of a GE1/1 ROB with twelve different strip configurations proposed for the M1 and M4 GE2/1 modules. These configuration explore different options for strip and trace lengths, as well as for strip and gap widths. To determine the geometry of the traces on the other side of the board, their length, width, and gap width are measured twelve times at each end of the traces. Because the gap width between the traces is not constant along their lengths, the average between the smallest and largest gap widths is used in the calculation. Table~\ref{tab:GE21_Trace} lists the specific configurations of the readout strips and the measured signal trace dimensions in each of the 12 sectors of the custom-built ROB, including the parameters from the designs in the original Technical Design Report (TDR) \cite{MuonTDR}.

\begin{figure}[!htp] 
\centering
\includegraphics[width=.40\columnwidth, height=0.54\textheight]{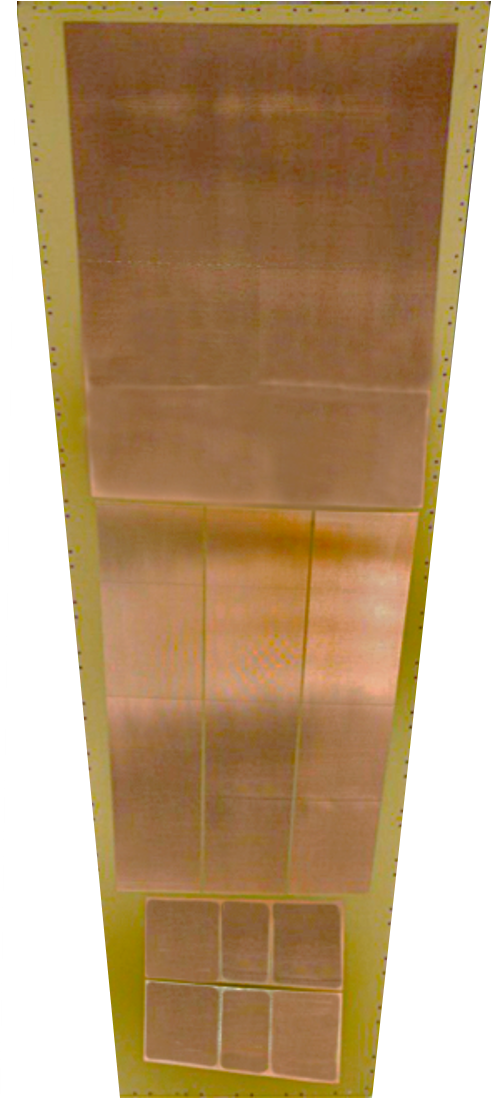}
\includegraphics[width=.40\columnwidth, height=0.54\textheight]{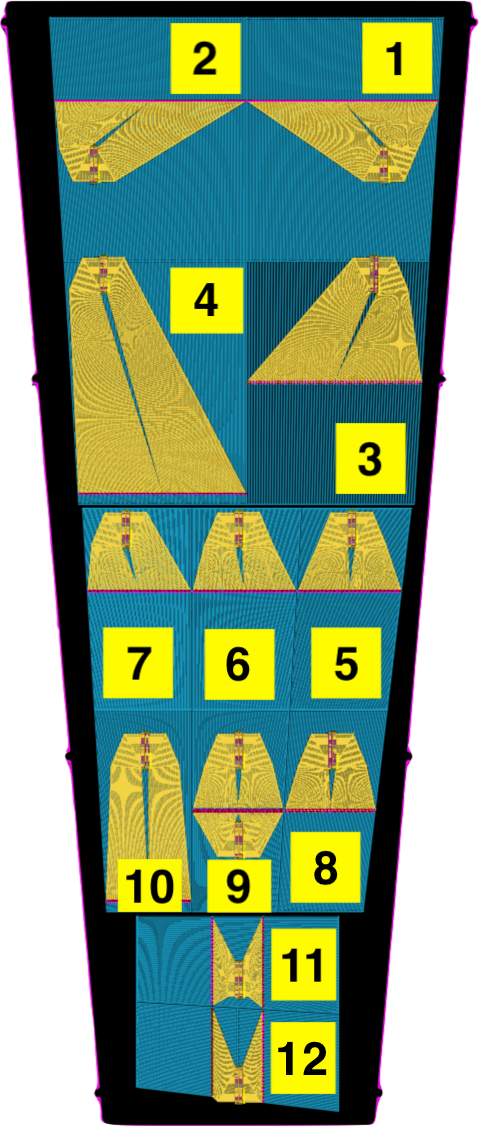}
\caption{\label{fig:GE21_ROB}Layout of the GE2/1 custom ROB. Left: Readout strips on the board on which interstrip capacitances are measured. Right: Signal trace routing on the back side of the ROB and the position of the readout connectors. The specific strip geometries in each of the 12 sectors are listed in table~\ref{tab:GE21_Trace}.}
\end{figure}

\begin{table}[!htp]
\centering
\caption{\label{tab:GE21_Trace}Trace dimensions for the ROB shown in figure~\ref{fig:GE21_ROB} (right).}
\smallskip
\resizebox{\textwidth}{!}{%
\begin{tabular}{|cccccc|}
\hline
Sector & Module & Parameters & Avg. meas. & Avg. meas. & Avg. meas. \\
       &        &            & length     & width      & Gap width\\
 &  &  & (cm) & (mm) & (mm)\\
\hline
1 & M4 & Original TDR design (Strip Gap: 0.2 mm) & 8.490 $\pm$ 1.796 & 0.48 $\pm$ 0.02 & 0.67 $\pm$ 0.02\\
2 & M4 & Gap: 0.3 mm & 9.234 $\pm$ 1.916 & 0.38 $\pm$ 0.02 & 0.72 $\pm$ 0.06\\
3 & M4 & 2$\times$Width, 0.5$\times$Length & 10.93 $\pm$ 1.59 & 0.44 $\pm$ 0.01 & 0.76 $\pm$ 0.05\\
4 & M4 & Long traces & 21.95 $\pm$ 0.91 & 0.38 $\pm$ 0.01 & 0.66 $\pm$ 0.02\\
5 & M1 & Original TDR design (Strip Gap: 0.2 mm) & 6.140 $\pm$ 1.122 & 0.60 $\pm$ 0.03 & 0.44 $\pm$ 0.03\\
6 & M1 & Gap: 0.3 mm & 5.438 $\pm$ 0.957 & 0.41 $\pm$ 0.01 & 0.59 $\pm$ 0.03\\
7 & M1 & Gap: 0.4 mm & 5.576 $\pm$ 1.027 & 0.39 $\pm$ 0.01 & 0.47 $\pm$ 0.03\\
8 & M1 & 2$\times$Width, 0.5$\times$Length & 6.209 $\pm$1.136 & 0.34 $\pm$ 0.01 & 0.45 $\pm$ 0.01\\
9 & M1 & 0.5$\times$Length & 1.990 $\pm$ 0.979 & 0.35 $\pm$ 0.02 & 0.51 $\pm$ 0.18\\
10 & M1 & Original TDR design, Long traces & 13.00 $\pm$ 0.07 & 0.33 $\pm$ 0.01 & 0.43 $\pm$ 0.01\\
11 & M1 & Original TDR design, Minimal traces & 3.425 $\pm$ 0.741 & 0.38 $\pm$ 0.01 & 0.40 $\pm$ 0.05\\
12 & M1 & Minimal traces, 2$\times$Width, 0.5$\times$Length & 3.806 $\pm$ 1.227 & 0.38 $\pm$ 0.01 & 0.45 $\pm$ 0.08\\
\hline
\end{tabular}%
}
\end{table}

The interstrip capacitance is measured in each readout sector with a commercial Excelvan M6013 capacitance meter. For each pair of strips, four measurements are made. To obtain an accurate measurement, the probes of the capacitance meter are placed at opposite ends of adjacent strip pairs, and held about a centimeter above the strips by one person while the meter is zeroed by another person. After zeroing, the probes are immediately placed on the strip and a reading is taken. A weighted mean over all strips in each sector is calculated from the average of all trials for each individual strip, and the standard deviation of the weighted mean is computed for all sectors.

\begin{figure}[!htp] 
\centering
\includegraphics[width=.9\textwidth, trim={0 10cm 0 7cm}, clip]{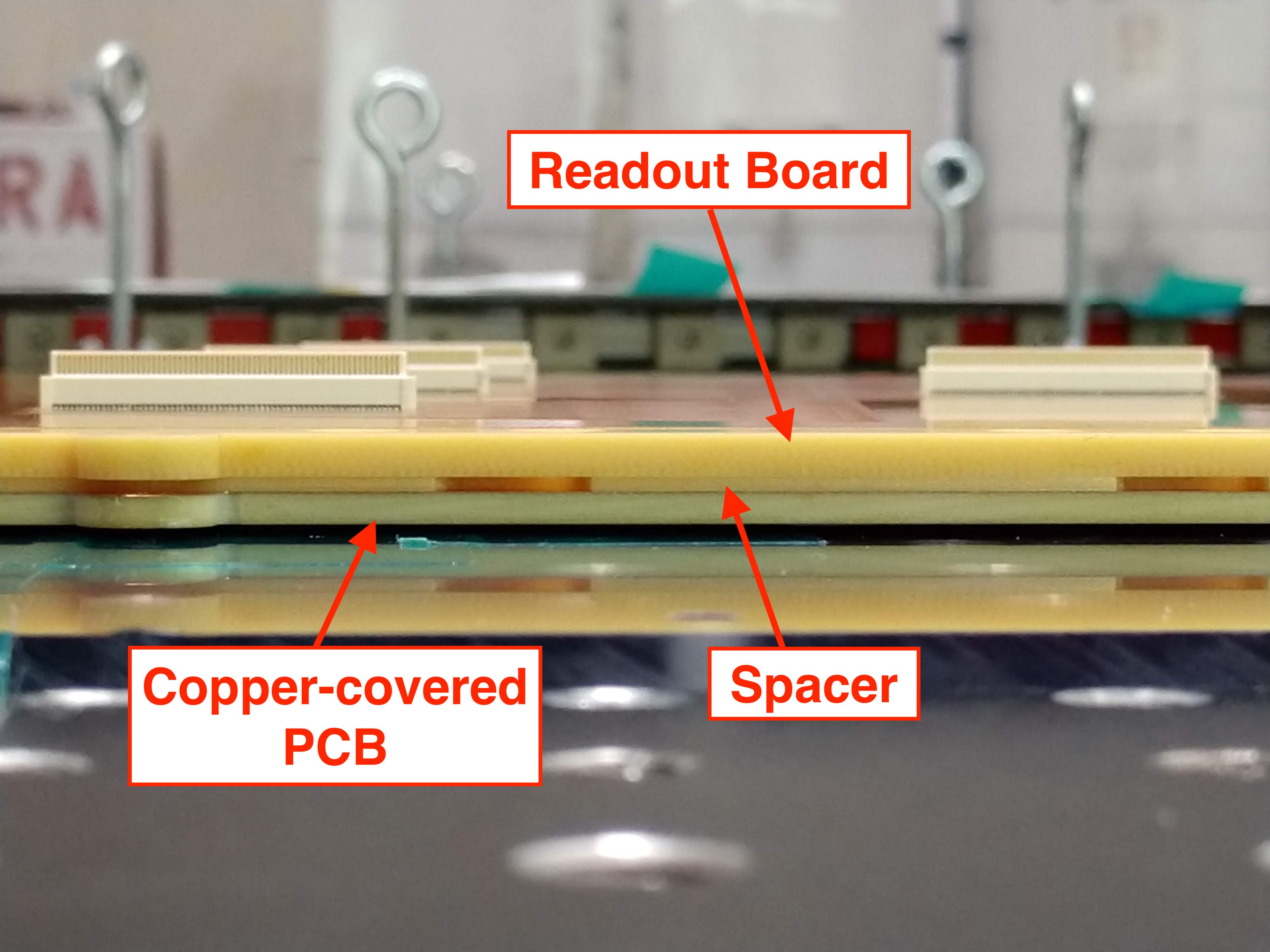}
\caption{\label{fig:ROBwPCB}Picture of a cross-section of the experimental configuration for the simulation of the capacitance contribution of the bottom of the third GEM foil. The readout board is suspended 1 mm above a copper-covered PCB with 1 mm FR4 spacers.}
\end{figure}

The measurements are repeated with a copper-covered PCB suspended 1 mm from the readout board in order to simulate the capacitance contribution of the bottom of the GEM3 foil above it (see figure \ref{fig:ROBwPCB}). One-millimeter dielectric FR4 spacers are placed around the edge of the ROB to hold the copper-clad PCB 1~mm from the ROB. Extra spacers are also used in areas where there are no readout strips to ensure that the ROB remains planar. Because the readout strips are not directly accessible in this configuration, the probes of the capacitance meter are instead placed on the pins of the 128-channel Panasonic connector on the ROB. The same procedure of zeroing the meter and reading the measurements is followed as described above.
The measurements are taken with the same statistics and in the same locations for both strips and traces (except for sector 1 where three additional measurements are taken).

\section{Results and discussion}
\label{sec:results}
  The calculated and measured interstrip capacitances for the readout board of GE2/1 detectors are presented for various configurations of strip dimensions. We wish to compare the results from the two calculation methods to each other and then the calculations to the measurements. For the latter, the trace dimensions in the analytical calculation and in the FEA modeling with 2 strips and 2 traces are varied to reproduce the different trace lengths on the physical ROB (see table~\ref{tab:GE21_Trace}).

\subsection{Comparison of results from analytical calculations and from FEA model}

\begin{figure}[!htb] 
\centering
\includegraphics[width=0.8\columnwidth, trim={0 0.3cm 0 1.0cm},clip=true]{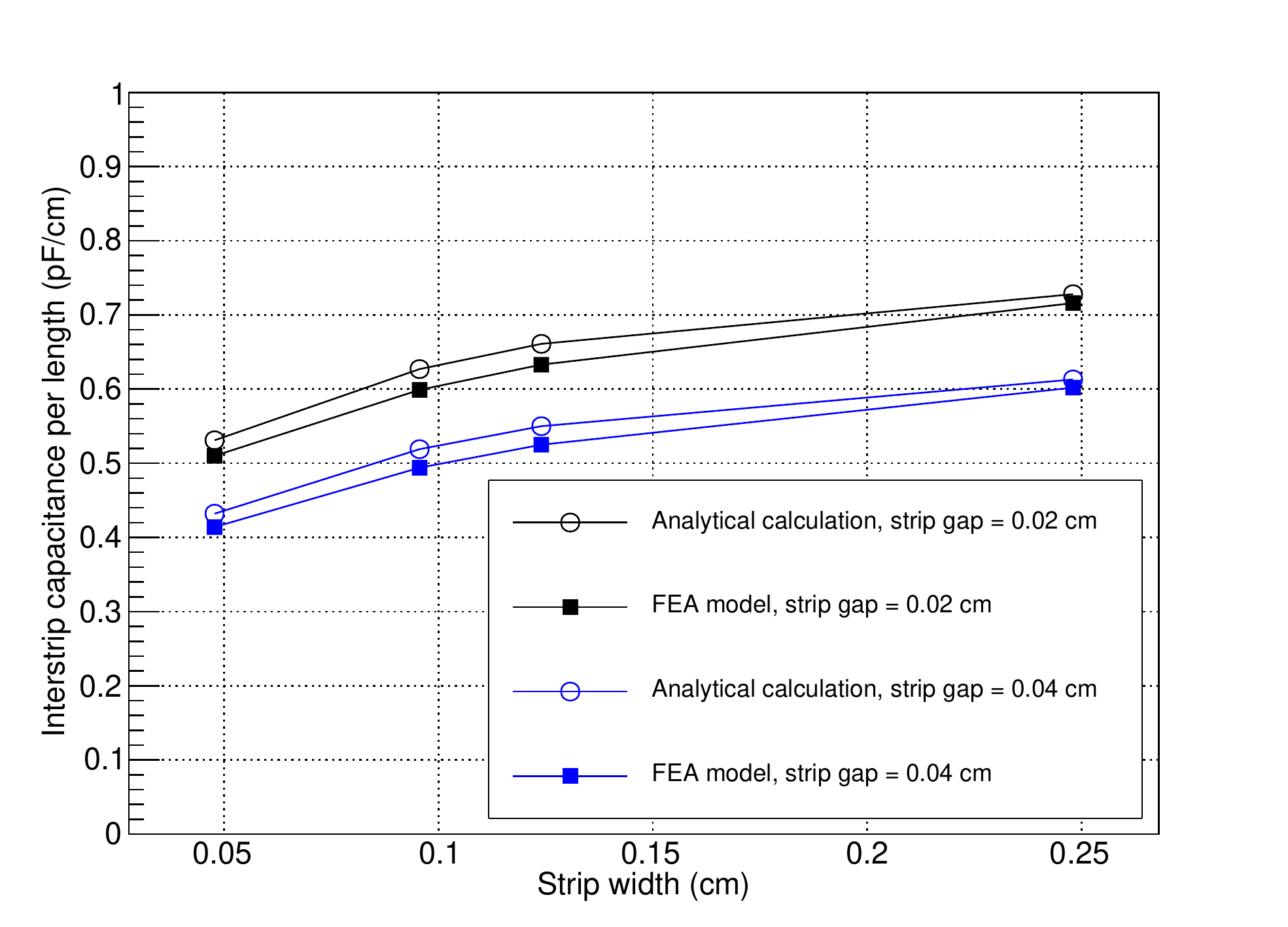}
\caption{\label{fig:Comp1}Comparison between analytical calculation and 2D FEA modeling of interstrip capacitances per length for original TDR strip width and doubled strip width in the M1 and M4 modules.}
\end{figure}

 Figure~\ref{fig:Comp1} compares the results obtained from varying the strip width and the gap width between the strips for the M1 and M4 modules. These results are based on the 2D model for both the analytical calculations and the FEA model. The four different strip widths considered here correspond to the original TDR strip width (see table~\ref{tab:GE21modules}) and a doubled strip width for the M1 and M4 modules. For a strip gap of 0.02~cm, the discrepancies between the analytical calculation and the FEA model are 1.5-4.5\%, and for a strip gap of 0.04 cm they are 1.8-4.8\%. For both strip gap configurations, the minimum error corresponds to the M4 module with double strip width and the maximum error is for the M1 module with double strip width.  
 
 Results from both methods agree quite well in all cases. As expected, interstrip capacitances increase when the strip width is doubled for both M1 and M4 modules. However, the increase is only on the order of 10-20\%. Since interstrip capacitance is directly proportional to strip length, halving the strip length will decrease the capacitance by 50\%. Consequently, halving the strip length while doubling the strip width does indeed lead to an overall reduction of the interstrip capacitance, which is now quantified to be 40-45\%. This result confirms the original premise for this study quantitatively. A decrease on the order of 20\% in the interstrip capacitance is seen in the calculations when the gap width between the strips is doubled.

\subsection{\label{sec:comp}Results from experimental measurements}

For the experimental measurements of the interstrip capacitance, the values listed in table~\ref{tab:GE21_all_woCu} are weighted means over all measured strip pairs in the sector, and the uncertainties are the standard deviations of the weighted means. The table also lists the ratios of the measured capacitances to the analytically calculated capacitances. 

\begin{table}[!htp]
\centering
\caption{\label{tab:GE21_all_woCu}Measured and calculated interstrip capacitances for open GE2/1 ROB.}
\smallskip
\resizebox{\textwidth}{!}{%
\begin{tabular}{|cccccc|}
\hline
Sector & Module & Parameters & Calc. cap. &  Avg. meas. cap. & \underline{Meas. cap.}\\
  &  &  & (pF) & (pF) & Calc. cap.\\
\hline
1 & M4 & Original TDR design (Strip Gap: 0.2 mm) & 16.7 &  21.69 $\pm$ 0.05 & 1.30\\
2 & M4 & Gap: 0.3 mm & 15.3 &  19.98 $\pm$ 0.12 & 1.31\\
3 & M4 & 2$\times$Width, 0.5$\times$Length & 10.5 &  15.32 $\pm$ 0.03 & 1.46\\
4 & M4 & Long traces & 21.1 & 27.87 $\pm$ 0.09 & 1.32\\
5 & M1 & Original TDR design (Strip Gap: 0.2 mm) & 12.7 & 16.27 $\pm$ 0.04 & 1.28\\
6 & M1 & Gap: 0.3 mm & 11.2 & 14.65 $\pm$ 0.07 & 1.31\\
7 & M1 & Gap: 0.4 mm & 10.6 & 13.17 $\pm$ 0.04 & 1.24\\
8 & M1 & 2$\times$Width, 0.5$\times$Length & 8.5 & 11.82 $\pm$ 0.06 & 1.39\\
9 & M1 & 0.5$\times$Length & 5.9 & 9.32 $\pm$ 0.05 & 1.58\\
10 & M1 & Original TDR design, Long traces & 15.3 & 20.07 $\pm$ 0.07 & 1.31\\
11 & M1 & Original TDR design, Minimal traces & 11.8 & 14.02 $\pm$ 0.02 & 1.19\\
12 & M1 & Minimal traces, 2$\times$Width, 0.5$\times$Length & 7.6 & 10.39 $\pm$ 0.07 & 1.37\\
\hline
\end{tabular}
}
\end{table}

The configuration with the lowest measured interstrip capacitance (9.32~$\pm$~0.05~pF) for the smaller M1 module is Sector 9 where the strip lengths are halved and the strip widths are unchanged, which is a 43\% reduction over the original TDR configuration (Sector 5). If in addition the strip widths are doubled as in Sector 8, the measured interstrip capacitance is 11.82~$\pm$~0.06~pF, which is a 27\% reduction over the original TDR configuration. For the larger M4 module, the configuration with the lowest measured interstrip capacitance (15.32 $\pm$ 0.03 pF) is Sector 3 where the strip lengths are halved and the strip widths are doubled. This is a 29$\%$ reduction over the original TDR configuration (Sector 1). These measured reductions trend in the same way as the corresponding calculated reductions (M1:\hspace{1mm}-33\%, M4:\hspace{1mm}-32\%) when the strip lengths are halved and the strip widths are doubled, but are slightly less pronounced.
Increasing the gap width decreases the interstrip capacitance as well (see results for Sectors 2, 6, and 7 in table~\ref{tab:GE21_all_woCu}) as one would expect, but the improvement over the original design is more marginal at the 8-16\% level. 

Table~\ref{tab:GE21_all_woCu} shows also that longer trace lengths increase the interstrip capacitance as one would expect. Comparing Sector 1 with Sector 4, i.e.\ the original configuration of the M4 module with the original configuration of the M4 module with long traces, we find that the interstrip capacitance is around 30$\%$ ($\sim$ 6~pF) larger for the latter in both the experimental measurements and the calculations. The same effect is observed when comparing Sector 5 with Sector 10, i.e.\ the original configuration of the M1 module with the original configuration of the M1 module but with long traces; here the measured interstrip capacitance is around 23$\%$ ($\sim$ 4~pF) larger for the latter. Similarly, comparing measurements in Sector 5 with Sector 11, i.e.\ the original configuration of the M1 module with the same geometrical configuration, but with minimal trace lengths, we find that the latter is around 14$\%$ ($\sim$ 2~pF) smaller. 

The measured capacitances are typically 20-40\% higher than the analytically calculated values for all configurations. This is due to the simplicity of the model that is applied in the calculations where only the two strips and the FR4 substrate are present. By contrast, on the ROB the measured strip pair is surrounded by many other strips which leads to a modification of the electric potential in the space around the strip pair and an increased capacitance. 

In summary, the measurements confirm the earlier conclusion from the calculations that the best strategy to minimize interstrip capacitance is to halve the strip lengths and double the strip widths if one wants to keep the number of strips constant. We also conclude that to reduce the interstrip capacitance and consequently the intrinsic noise from the ROB, the length of the traces should be minimized as much as possible in the GE2/1 and ME0 ROB designs.
This is exemplified by the measurement result for Sector 12 that shows a 36\% reduction in interstrip capacitance when both strategies are employed together.

\subsection{Results for interstrip capacitance with and without a copper-covered PCB and 3D model}

The motivation to perform additional measurements with the presence of a conductor plane comes from the discrepancy observed between the analytical calculations and the measurements shown in table~\ref{tab:GE21_all_woCu}. Clearly, nearby conductors modify the capacitance between adjacent strips. In a CMS GEM detector, the bottom of GEM3 represents a large additional electrode only 1~mm away from the strips.
The average measured interstrip capacitances both with and without the presence of a copper-covered PCB to simulate the bottom of GEM3, and the ratio of the two, are presented for each sector of the ROB in table~\ref{tab:GE21_Meas}. With the PCB present, the measured capacitances change by -2\% to +34\% over the measurements without PCB with most sectors showing an increased capacitance with an average of +15\%.

The basic conclusions from the previous section still hold in this more realistic configuration. The measured interstrip capacitances with halved strip lengths and doubled strip widths are reduced by 17\% and 22\% over the original TDR configurations for the M1 and M4 modules, respectively. Sector 12 shows a 46\% reduction in interstrip capacitance over the original TDR configuration (Sector 5) when in addition the trace lengths are minimized in M1.

\begin{table}[!htp]
\centering
\caption{\label{tab:GE21_Meas}Measurements of the GE2/1 ROB interstrip capacitance with and without a facing copper plate.}
\smallskip
\resizebox{\textwidth}{!}{%
\begin{tabular}{|cccccc|}
\hline
Sector & Module & Parameters & Avg. meas. cap. & Avg. meas. cap. & $C_{w}/C_{w/o}$\\
 &  &  & w/o plate & with plate & \\
  &  &  & (pF) & (pF) & \\
\hline
1 & M4 & Original TDR design (Strip Gap: 0.2 mm) & 21.69 $\pm$ 0.05 & 25.85 $\pm$ 0.27 & 1.192 $\pm$ 0.013\\
2 & M4 & Gap: 0.3 mm & 19.98 $\pm$ 0.12  & 20.98 $\pm$ 0.03 & 1.050 $\pm$ 0.006\\
3 & M4 & 2$\times$Width, 0.5$\times$Length & 15.32 $\pm$ 0.03 & 20.16 $\pm$ 0.04 & 1.315 $\pm$ 0.004\\
4 & M4 & Long traces & 27.87 $\pm$ 0.09 & 28.43 $\pm$ 0.07 & 1.020 $\pm$ 0.003\\
5 & M1 & Original TDR design (Strip Gap: 0.2 mm) & 16.27 $\pm$ 0.04 & 19.04 $\pm$ 0.21 & 1.170 $\pm$ 0.013\\
6 & M1 & Gap: 0.3 mm & 14.65 $\pm$ 0.07 & 18.26 $\pm$ 0.06 & 1.246 $\pm$ 0.007\\
7 & M1 & Gap: 0.4 mm & 13.17 $\pm$ 0.04 & 14.85 $\pm$ 0.08 & 1.128 $\pm$ 0.007\\
8 & M1 & 2$\times$Width, 0.5$\times$Length & 11.82 $\pm$ 0.06 & 15.82 $\pm$ 0.32 & 1.338 $\pm$ 0.028\\
9 & M1 & 0.5$\times$Length & 9.32 $\pm$ 0.05 & 9.17 $\pm$ 0.10 & 0.984 $\pm$ 0.012\\
10 & M1 & Original TDR design, Long traces & 20.58 $\pm$ 0.06 & 26.80 $\pm$ 0.14 & 1.302 $\pm$ 0.008\\
11 & M1 & Original TDR design, Minimal traces & 14.02 $\pm$ 0.02 & 14.82 $\pm$ 0.03 & 1.057 $\pm$ 0.003\\
12 & M1 & Minimal traces, 2$\times$Width, 0.5$\times$Length & 10.39 $\pm$ 0.07 & 10.33 $\pm$ 0.11 & 0.994 $\pm$ 0.013\\
\hline
\end{tabular}%
}
\end{table}

The interstrip capacitance between two adjacent strips obtained from the multi-strip FEA model both with and without the presence of a copper plate, and the ratio of the two are presented in table~\ref{tab:GE21_FEA}. Because the FEA model considers multiple strips, it is difficult to simulate in addition traces of varying lengths and widths. Consequently, in this model the traces are not considered, but the various strip and gap geometries are simulated, which results in simulations performed for eight of the twelve ROB sectors. The interstrip capacitances with and without a copper plate are almost the same in the simulation, as can be seen from the ratio between them. As in the experimental measurements, the lowest simulated interstrip capacitance for the M1 module is obtained by halving the strip length, and for the M4 module by doubling the strip width and halving the strip length.

Although the overall relative variations of the FEA results from sector to sector are in agreement with the variations in the experimental measurements, there exist differences in the absolute values. Comparing table~\ref{tab:GE21_Meas} with table~\ref{tab:GE21_FEA}, we find that the values obtained from the FEA model are 22-40\% lower than the experimentally measured values. This discrepancy is presumably due to the absence of the signal traces when simulating the readout board using the FEA model.

\begin{table}[!htp]
\centering
\caption{\label{tab:GE21_FEA}Interstrip capacitance from multi-strip FEA model of GE2/1 ROB with and without a copper plate. Note that sectors 4, 10-12 that vary trace lengths are not considered in the model and the error in the values of interstrip capacitance is $10^{-3}$~pF.}
\smallskip
\resizebox{\columnwidth}{!}{%
\begin{tabular}{|cccccc|}
\hline
Sector & Module & Parameters & FEA cap. & FEA cap. & $C_{w}/C_{w/o}$\\
 &  &  & w/o plate & with plate & \\
  &  &  & (pF) & (pF) & \\
\hline
1 & M4 & Original TDR design (Strip Gap: 0.2 mm) & 16.12 & 16.25 & 1.008\\
2 & M4 & Gap: 0.3 mm & 14.93 & 14.97 & 1.003\\
3 & M4 & 2$\times$Width, 0.5$\times$Length & 12.64 & 12.69 & 1.004\\
4 & M4 & Same as Sector 1 & 16.12 & 16.25 & 1.008\\
5 & M1 & Original TDR design (Strip Gap: 0.2 mm) & 13.16 & 13.61 & 1.034\\
6 & M1 & Gap: 0.3 mm & 11.24 & 11.27 & 1.002\\
7 & M1 & Gap: 0.4 mm & 10.43 & 10.56 & 1.013\\
8 & M1 & 2$\times$Width, 0.5$\times$Length & 10.54 & 10.74 & 1.018\\
9 & M1 & 0.5$\times$Length & 7.01 & 7.18 & 1.024\\
10 & M1 & Same as Sector 5 & 13.16 & 13.61 & 1.034\\
11 & M1 & Same as Sector 5 & 13.16 & 13.61 & 1.034\\
12 & M1 & Same as Sector 8 & 10.54 & 10.74 & 1.018\\
\hline
\end{tabular}
}
\end{table}

\begin{figure}[!htp] 
\centering
\includegraphics[width=0.49\columnwidth]{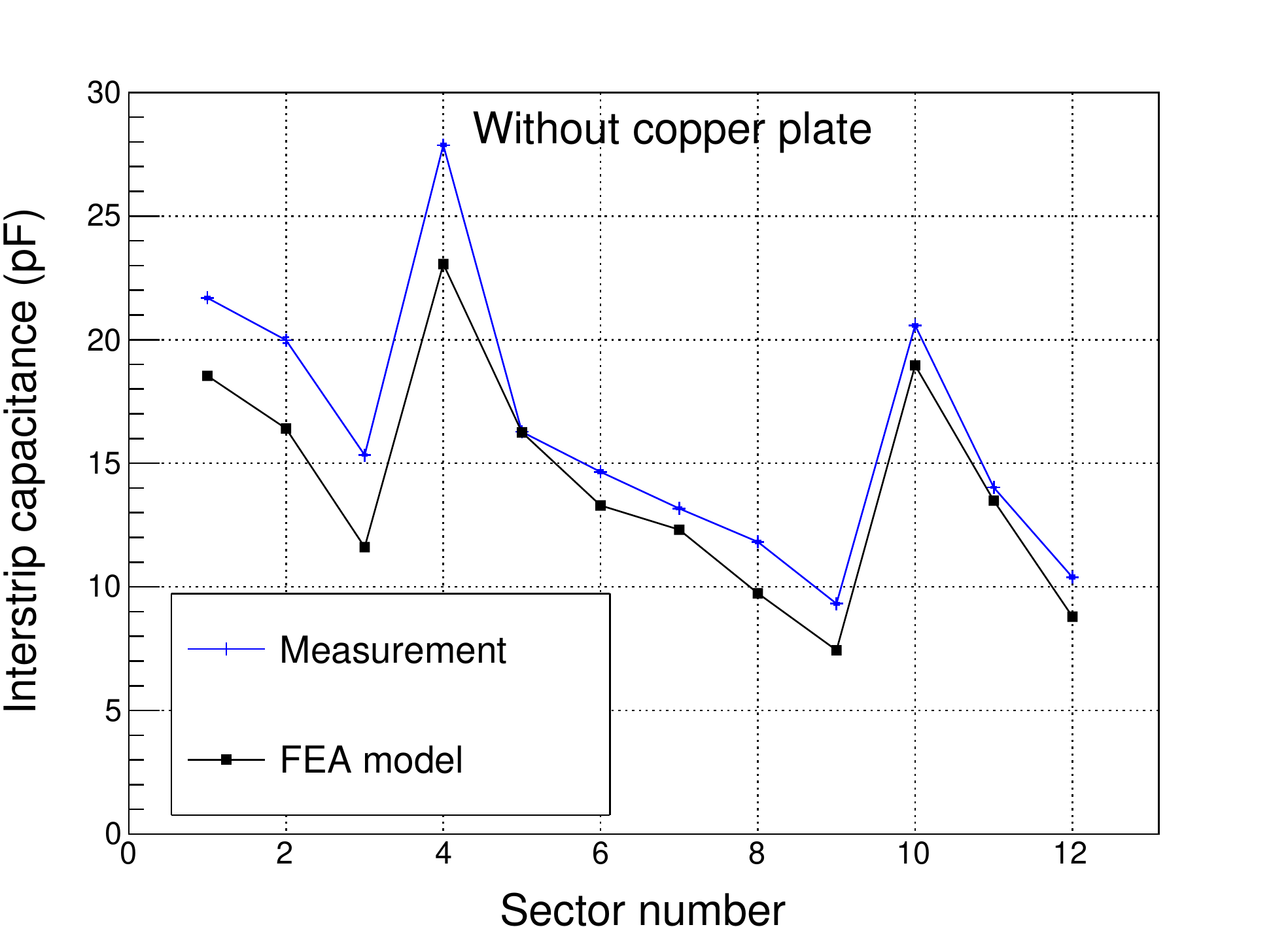}
\includegraphics[width=0.49\columnwidth]{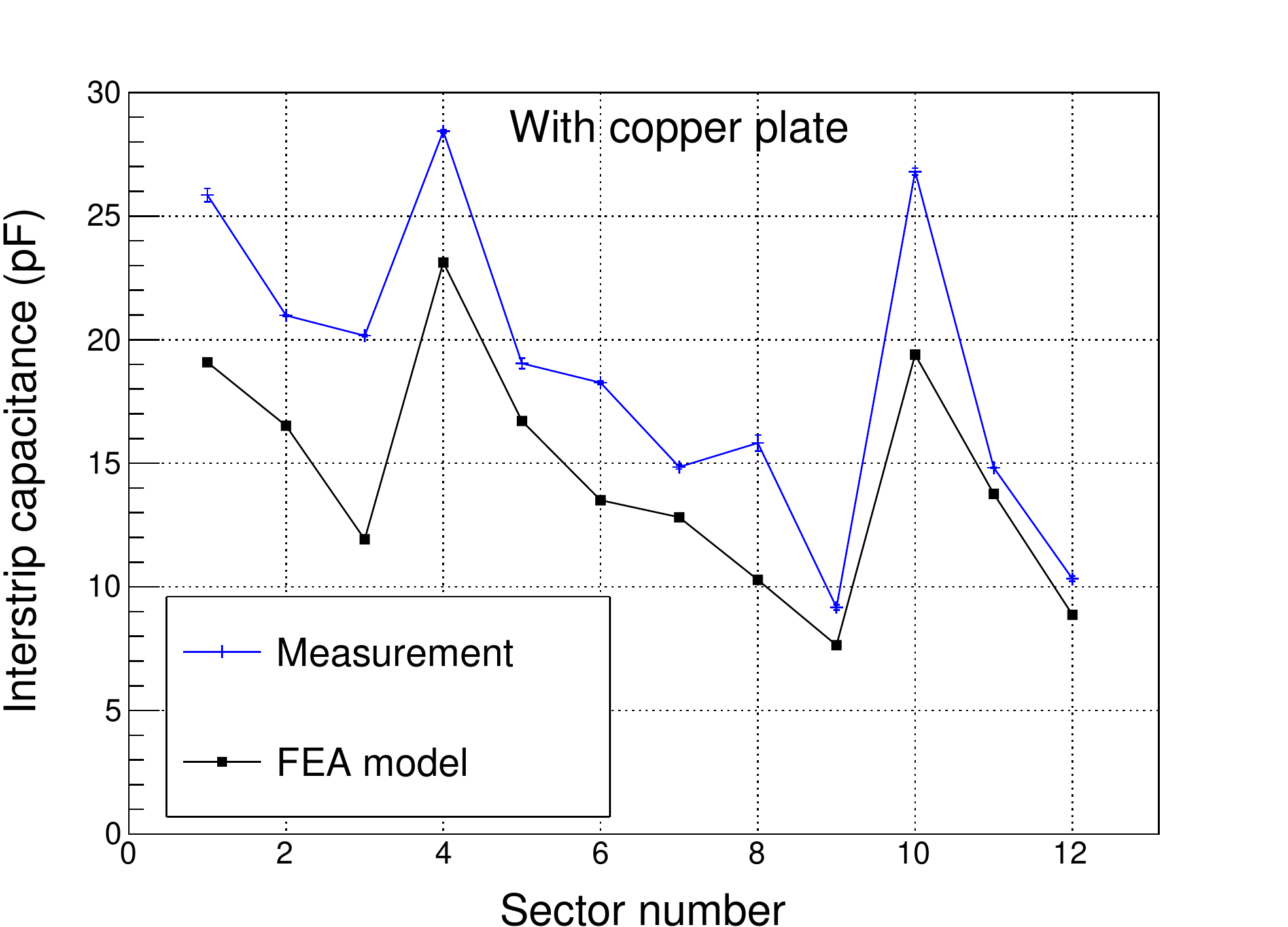}
\caption{\label{fig:Comp2}Left: Comparison between experimental measurements and results from two-strips-and-two-traces FEA model in 3D without a facing copper plate. Right: Comparison between experimental measurements and results from two-strips-and-two-traces FEA model in 3D with a facing copper plate. Note: The error bars for the measurements are present but are of the same order as the markers.}
\end{figure}

This observed discrepancy prompts us to create a 3D FEA model to include the contribution of the traces. However, it is technically difficult to simulate the traces of the whole custom-built GE2/1 readout board (see figure~\ref{fig:GE21_ROB}), so a simple model with only two strips and two traces is considered to model the 12 sectors with their various strip and trace geometries (see figure~\ref{fig:strips_traces}). This model can also simulate the effect of the presence of copper at the bottom of the GEM3 foil. Table~\ref{tab:GE21_FEA_2} shows the results of the interstrip capacitance obtained from this two-strips-and-two-traces 3D FEA model, both with and without the presence of a copper plate, and the ratio of the two. Including the traces in the model reduces the relative discrepancies with the experimental measurements significantly. Figure~\ref{fig:Comp2} shows a comparison of the results from this FEA model and experimental measurements, both without a copper plate (left) and with a copper plate (right). Results without copper plate typically agree better than results with copper plate.

\begin{table}[!htp]
\centering
\caption{\label{tab:GE21_FEA_2}Interstrip capacitance from two-strips-and-two-traces 3D FEA model of GE2/1 ROB with and without a copper plate. Note: The error in the values of interstrip capacitance is $10^{-3}$pF.}
\smallskip
\resizebox{\textwidth}{!}{%
\begin{tabular}{|cccccc|}
\hline
Sector & Module & Parameters & FEA cap. & FEA cap. & $C_{w}/C_{w/o}$\\
 &  &  & w/o plate & with plate & \\
  &  &  & (pF) & (pF) & \\
\hline
1 & M4 & Original TDR design (Strip Gap: 0.2 mm) & 18.538 & 19.085 & 1.029\\
2 & M4 & Gap: 0.3 mm & 16.405  & 16.519 & 1.007\\
3 & M4 & 2$\times$Width, 0.5$\times$Length & 11.606 & 11.924 & 1.027\\
4 & M4 & Long traces & 23.060 & 23.124 & 1.002\\
5 & M1 & Original TDR design (Strip Gap: 0.2 mm) & 16.251 & 16.709 & 1.028\\
6 & M1 & Gap: 0.3 mm & 13.294 & 13.506 & 1.015\\
7 & M1 & Gap: 0.4 mm & 12.313 & 12.816 & 1.041\\
8 & M1 & 2$\times$Width, 0.5$\times$Length & 9.741 & 10.283 & 1.055\\
9 & M1 & 0.5$\times$Length & 7.435 & 7.639 & 1.027\\
10 & M1 & Original TDR design, Long traces & 18.960 & 19.394 & 1.022\\
11 & M1 & Original TDR design, Minimal traces & 13.489 & 13.764 & 1.020\\
12 & M1 & Minimal traces, 2$\times$Width, 0.5$\times$Length & 8.796 & 8.873 & 1.008\\
\hline
\end{tabular}%
}
\end{table}

\begin{figure}[!htp] 
\centering
\includegraphics[width=0.8\columnwidth]{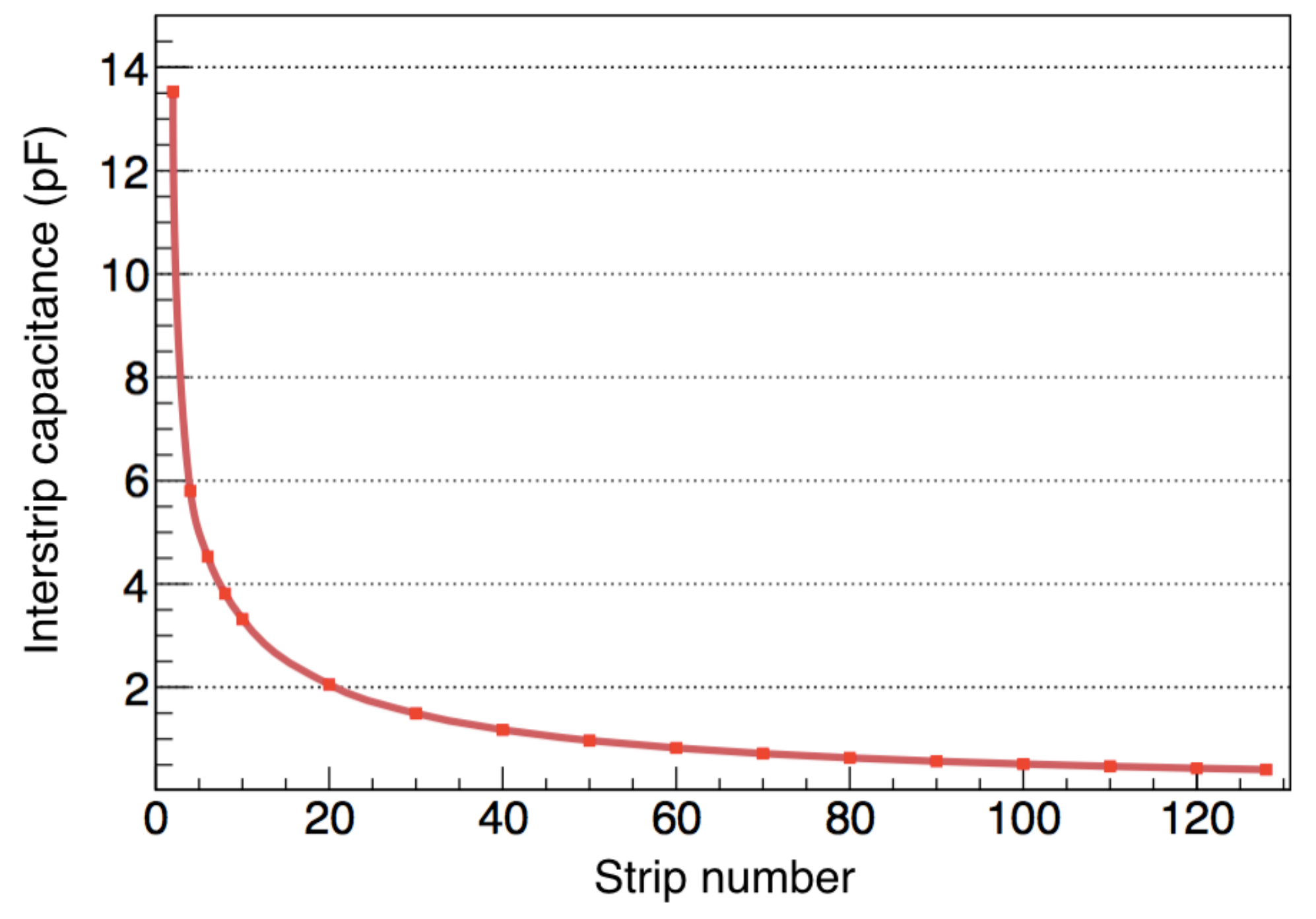}
\caption{\label{fig:IC_var}Interstrip capacitance between first strip and a non-adjacent strip using the multi-strip FEA model. For this calculation, a fixed potential of 1~V is applied to the first strip and a fixed potential of 0~V is applied to the other strip of interest.}
\end{figure}

The multi-strip FEA model allows us also to calculate an interstrip capacitance between strips that are not necessarily directly adjacent. For example, we calculate the capacitance between the first strip and each of the next 128 strips in the presence of all the other strips as shown in figure~\ref{fig:3_128_strips} (right). A fixed potential of 1~V is applied to the first strip and a fixed potential of 0~V is applied to the other strip of interest. The capacitance between those two strips is then calculated with the potential of all other strips varying accordingly. The result is shown in figure~\ref{fig:IC_var} for the first 128 strips without a PCB present. Beyond that, the capacitance does not change very much anymore. As the distance between the strips is increased, the interstrip capacitance decreases similar to an inverse function.

\section{Summary and conclusions}
This paper presents and discusses analytical calculations and physical measurements of the interstrip capacitances for twelve different potential strip geometries and dimensions of the readout boards for the smallest (M1) and largest (M4) modules in the GE2/1 detector for the CMS muon upgrade. We also present results from 2D and 3D Finite Element Analysis modeling of this system. The main goal of the study is to find configurations that minimize the interstrip capacitances and consequently maximize the signal-to-noise ratio for the detector. Specifically, we investigate if a configuration with doubled strip width and halved strip length, which leaves total channel number in the detector unchanged, can reduce interstrip capacitance compared with the original configuration.

Overall, we find agreement at the 1.5--4.8\% level between the two methods of calculations and on the average at the 17\% level between calculations and measurements. For the M1 (M4) module, the configuration with halved strip lengths and doubled strip widths results in a measured 27 (29)$\%$ reduction over the original configuration. The corresponding calculations give reductions of 33 (32)\%. Increasing the width of the 0.02~cm gaps  between strips by 50--100\% only produces a 8--16\% reduction in interstrip capacitance.
An important observation from the measurements of the interstrip capacitance is the effect the signal traces on the other side of the board have on the capacitance. With longer trace lengths, the interstrip capacitance increases by about 23 (30)\% in the M1 (M4) module. We also find on average a 15\% increase in the measured interstrip capacitance with the addition of a copper-covered PCB which is used to simulate the capacitance contribution from the bottom of the third GEM foil. 

Finally, we briefly comment on the expected impact of the doubled strip width on detector performance. The main purpose of the GE2/1 detector is to help control the Level-1 muon trigger rates. In the original design~\cite{CMSTDR,MuonTDR}, two strips are ganged together to form a trigger "pad". Consequently, we expect no impact on trigger performance at all if one double-width strip is used as one trigger "pad". 


We conclude that the best strategy to minimize interstrip capacitance for the GE2/1 readout boards is indeed to halve the strip lengths and double the strip widths if one wants to keep the number of strips constant and to minimize the length of all signal traces on the board. We have now adopted this design modification for all eight module types of the GE2/1 detector and will produce the final detector with this new strip design (see figure~\ref{fig:M5ROB}).


\section*{Acknowledgments}
We gratefully acknowledge support from FRS-FNRS (Belgium), FWO-Flanders (Belgium), BSF-MES (Bulgaria), MOST and NSFC (China), BMBF (Germany), CSIR (India), DAE (India), DST (India), UGC (India), INFN (Italy), NRF (Korea), QNRF (Qatar), DOE (USA), and NSF (USA). 
We thank Dr. Shady Khalil for providing access to COMSOL Multiphysics software and Qasim Khan for the initial setup of the package for this work. Both are from the Electrical Engineering Department of Texas A\&M University at Qatar. The authors also wish to thank a team of undergraduate students at Florida Institute of Technology (A. Bustos, J. Hammond, A. Lucciola, M. Werbiskis, and L. Shaw), for their help with performing the interstrip capacitance and trace dimension measurements.



\begin{thebibliography}{99}
\bibitem{CMS} CMS Collaboration, \textit{The CMS experiment at the CERN LHC}, \href{https://doi.org/10.1088%2F1748-0221%2F3%2F08%2Fs08004}{\textit{JINST} \textbf{3} (2008) S08004}.

\bibitem{HiggsATLAS} ATLAS Collaboration, \textit{Observation of a new particle in the search for the standard model Higgs boson with the ATLAS detector at the LHC}, \href{https://doi.org/10.1016/j.physletb.2012.08.020}{\textit{Phys. Lett. B} \textbf{716} (2012) 1}. 

\bibitem{HiggsCMS} CMS Collaboration, \textit{Observation of a new boson at a mass of 125 GeV with the CMS experiment at the LHC}, \href{https://doi.org/10.1016/j.physletb.2012.08.021}{\textit{Phys. Lett. B} \textbf{716} (2012) 30}.
 
\bibitem{CMSTDR} CMS Collaboration, \textit{Technical Proposal for the Phase-II Upgrade of the Compact Muon Solenoid, Technical Report} \href{https://cds.cern.ch/record/2020886}{\textbf{CERN-LHCC-2015-010}, \textbf{CMS-TDR-15-02} (2015)}. 

\bibitem{MuonTDR} CMS Collaboration, \textit{The Phase-2 Upgrade of the CMS Muon Detectors, Technical Report} \href{https://cds.cern.ch/record/2283189}{\textbf{CERN-LHCC-2017-012}, \textbf{CMS-TDR-016} (2017)}. 

\bibitem{Sauli} F. Sauli, \textit{GEM: A new concept for electron amplification in gas detectors}, \href{https://doi.org/10.1016/S0168-9002(96)01172-2}{\textit{Nucl. Instrum. Meth. A} \textbf{386} (1997) 531}. 

\bibitem{GEMTDR} CMS Collaboration, \textit{CMS Technical Design Report for the Muon Endcap GEM Upgrade, Technical Report} \href{https://cds.cern.ch/record/2021453}{\textbf{CERN-LHCC-2015-012}, \textbf{CMS-TDR-013} (2015)}.

\bibitem{Capa1} S. Gevorgian and H. Berg, \textit{Line Capacitance and Impedance of Coplanar-Strip Waveguides on Substrates with Multiple Dielectric Layers} in  \href{https://doi.org/10.1109/EUMA.2001.339161}{\textit{Proc.\ 31$^{st}$ European Microwave Conference}}, London, England, 2001, pp. 1-4.

\bibitem{Capa2} J. M. Martinis, R. Barends, and A. N. Korotkov, \textit{Calculation of Coupling Capacitance in Planar Electrodes}, [\href{https://arxiv.org/abs/1410.3458}{arXiv:1410.3458}] (2014). 

\bibitem{MATLAB} The MathWorks Inc., MATLAB, version R2018a, \href{https://www.mathworks.com/products/matlab.html}{https://www.mathworks.com/products/matlab.html}, Natick, Massachusetts, USA, 2018.

\bibitem{COMSOL} COMSOL Inc., COMSOL Multiphysics®, version 5.4, \href{https://www.comsol.com}{www.comsol.com}, COMSOL AB, Stockholm, Sweden, 2018.

\end{thebibliography}
\end{document}